\documentclass[aps,prd,twocolumn,nofootinbib,showpacs,superscriptaddress,floatfix]{revtex4-1}

\usepackage{amsmath}
\usepackage{epsfig}
\usepackage{graphicx}
\usepackage{subfigure}

\newcommand\be{\begin{equation}}
\newcommand\ba{\begin{eqnarray}}
\newcommand\ee{\end{equation}}
\newcommand\ea{\end{eqnarray}}

\newcommand\bw{\begin{widetext}}
\newcommand\ew{\end{widetext}}

\newcommand{\nn}{\nonumber}

\newcommand{\Hor}{{\mbox{\tiny Hor}}}
\newcommand{\MGBK}{{\mbox{\tiny MGBK}}}
\newcommand{\QK}{{\mbox{\tiny QK}}}
\newcommand{\BK}{{\mbox{\tiny BK}}}
\newcommand{\MN}{{\mbox{\tiny MN}}}
\newcommand{\MK}{{\mbox{\tiny MK}}}

\newcommand{\Kerr}{{\mbox{\tiny K}}}

\begin{document}

\title{Systematic study of event horizons and pathologies of parametrically deformed Kerr spacetimes}

\author{Tim Johannsen}
\affiliation{Department of Physics and Astronomy, University of Waterloo, Waterloo, Ontario, N2L 3G1, Canada}
\affiliation{Canadian Institute for Theoretical Astrophysics, University of Toronto, Toronto, Ontario M5S 3H8, Canada}
\affiliation{Perimeter Institute for Theoretical Physics, Waterloo, Ontario, N2L 2Y5, Canada}
\affiliation{Physics Department, University of Arizona, Tucson, Arizona 85721, USA}

\date{\today}

\begin{abstract}

In general relativity, all black holes in vacuum are described by the Kerr metric, which has only two independent parameters: the mass and the spin. The unique dependence on these two parameters is known as the ``no-hair'' theorem. This theorem may be tested observationally by using electromagnetic or gravitational-wave observations to map the spacetime around a candidate black hole and measure potential deviations from the Kerr metric. Several parametric frameworks have been constructed for tests of the no-hair theorem. Due to the uniqueness of the Kerr metric, any such parametric framework must violate at least one of the assumptions of the no-hair theorem. This can lead to pathologies in the spacetime, such as closed timelike curves or singularities, which may hamper using the metric in the strong-field regime. In this paper, I analyze in detail several parametric frameworks and show explicitly the manner in which they differ from the Kerr metric. I calculate the coordinate locations of event horizons in these metrics, if any exist, using methods adapted from the numerical relativity literature. I identify the regions where each parametric deviation is unphysical as well as the range of coordinates and parameters for which each spacetime remains a regular extension of the Kerr metric and is, therefore, suitable for observational tests of the no-hair theorem.

\end{abstract}



\maketitle

\section{Introduction}

According to the no-hair theorem, the exterior gravitational fields of isolated and stationary black holes in general relativity are uniquely characterized by their mass $M$ and (the magnitude of their) spin angular momentum $J$. In particular, this field is described by the Kerr metric, the unique stationary, axisymmetric, asymptotically flat, vacuum solution to the Einstein equations that possesses an event horizon and is free of closed timelike curves outside of the horizon \cite{Penrose69,NHT,rigidity,Carter73,Price72}. The no-hair theorem, therefore, implies that all multipole moments of the Kerr spacetime are completely determined by the first two, the mass monopole and spin dipole. When these multipole moments are written as a set of mass multipole moments $M_l$, which are nonzero for even values of $l$, and a set of current multipole moments $S_l$, which are nonzero for odd values of $l$, the no-hair theorem is captured by the relation \cite{Geroch70}
\begin{equation}
M_{l}+{\rm i}S_{l}=M({\rm i}a)^{l},
\label{kerrmult}
\end{equation}
where $a\equiv J/M$ is the spin parameter. 

Thanks to the no-hair theorem, this property of black holes in general relativity naturally leads to the expectation that all astrophysical black holes are described by the Kerr metric. Astrophysical black holes, however, will be neither perfectly stationary, nor exist in perfect vacuum. Other stars, electromagnetic fields, and other forms of matter like dust and dark matter, will induce perturbations away from the Kerr metric. Such perturbations will induce nonzero deviations from Eq.~\eqref{kerrmult} that could lead to a violation of the no-hair theorem. However, if one makes the implicit assumption that such perturbations will be so small to be essentially unobservable, then one can argue that astrophysical black holes must be described by the Kerr metric. This is the assumption I make in this paper.    

There exists observational evidence for the presence of horizons in astrophysical black holes (see the discussion in, e.g., Ref.~\cite{Narayan}), but a proof of the validity of the no-hair theorem is still lacking. This is why a concrete effort has been brewing for the past few years to develop model-independent tests using electromagnetic and gravitational-wave observations to determine the precise strong-field nature of black holes. Instead of focusing on particular gravity theories and introducing modifications of the Einstein-Hilbert action, these tests are designed in a phenomenological approach that encompasses large classes of modified theories of gravity and which is able to test many different theories simultaneously. In this case, the underlying fundamental theory is usually unknown, and insight is hoped to be gained primarily through observations.

Such strong-field tests can be classified into two groups (see Ref.~\cite{reviews} for reviews on this topic): 
\begin{itemize}
\item Gravitational-wave tests using the gravitational waves generated by stellar-mass compact objects in tight extreme-mass ratio inspiral (EMRI) orbits around  a supermassive black hole~\cite{Ryan95,EMRI,kludge,CH04,GB06,Brink08,Gair08,Apostolatos09,VH10,Vigeland:2010xe,Vigeland:2011ji,Gair:2011ym}; 
\item Electromagnetic tests using the radiation emitted by accelerating particles in an accretion disk around a black hole \cite{PaperI,PaperII,PaperIII,EM,Bambi2012,BambiBarausse,PJ11,PaperIV}.
\end{itemize}
Other weak-field tests of the no-hair theorem exist, such as those obtained from observing close stellar orbits around Sgr~A*~\cite{Sgr} and pulsar/black-hole binaries~\cite{WK99}, but these do not probe the near-horizon, strong-field nature of black holes. While extensive searches of pulsars orbiting around black holes are ongoing (e.g., Ref.~\cite{Smits}), no pulsar/black-hole binary has been found so far.

In contrast to weak-field tests of gravity, where a parameterized post-Newtonian approach (e.g.,~\cite{Will93}) is sufficient, strong-field tests require a careful modeling of the spacetime by introducing a parametric deviation from the Kerr metric. Several such parametric frameworks have been constructed, within which possible observational signatures of deviations from the Kerr metric can be explored (e.g.,~\cite{MN92,CH04,GB06,VH10,Vigeland11,MKmetric}). The objects they describe have spacetimes that can deviate slightly to severely from the Kerr metric, and observables can be studied in terms of one or more free parameters. All of these metrics contain the Kerr metric as the special case, when the deviations are dialed to zero.

The many proposed metrics in the literature can be divided into two subclasses: those that are Ricci flat, i.e., $R_{\mu \nu} = 0$, and those that are not. In the former case, the metric in the far field satisfies the Laplace equation, and thus, when in asymptotically Cartesian and mass-centered coordinates, it can be expressed as a sum of mass and current multipole moments (see, e.g., Ref.~\cite{Thorne:1980ru}). One can relate these moments to each other via~\cite{CH04,VH10,Vigeland:2010xe}
\begin{equation}
M_{\ell} + {\rm i}S_{\ell} = M({\rm i}a)^{{\ell}} + \delta M_{\ell} + {\rm i}\delta S_{\ell}\,,
\label{mult}
\end{equation}
where $\delta M_{\ell}$ and $\delta S_{\ell}$ are mass and current multipole deformations. For this class of metrics, the measurements of three or more multipole moments could then be used to test for deviations in the Kerr metric~\cite{Ryan95}. 

When the metric is not Ricci flat, the above parameterization of the metric in the far field (as a sum over mass and current multipole moments that depend only on the $\ell$ harmonic number) is not valid. Such metrics generically arise from explicit or implicit modifications to the Einstein-Hilbert action, such as in dynamical Chern-Simons gravity~\cite{YP09,SY09,Alexander:2009tp} and in Einstein-dilaton-Gauss-Bonnet gravity~\cite{Yunes:2011we}. In these cases, it is not clear what the general structure of a modification of Eq.~\eqref{kerrmult} would look like.

If a gravitational wave or electromagnetic measurement requires a Kerr deviation, then there are two possible implications. One possibility is that the object observed is not an ideal astrophysical black hole. This could mean that either the black hole is not perfectly stationary or it is not in pure vacuum, indicating a failure of my assumption that these prosaic deviations are unobservable. More interestingly, the object may not be a black hole at all, but is instead a more exotic object~\cite{CH04,Hughes06} that perhaps violates cosmic censorship~\cite{Penrose69}. Another possibility is that four-dimensional general relativity is only approximately valid in the strong-field regime, and thus, stationary and vacuum black holes solutions are not described by the Kerr metric (see, e.g., Refs.~\cite{YP09,SY09,Psaltis08,Yunes:2011we}). In this interpretation, if the object is otherwise known to possess a horizon, both the no-hair theorem and strong-field general relativity are invalid.

In this paper, I analyze the properties of several parametric deviations from the Kerr spacetime. I consider the quasi-Kerr metric of Glampedakis and Babak (QK; \cite{GB06}), the bumpy Kerr metric of Vigeland and Hughes (BK; \cite{CH04,VH10}), the metric proposed by Manko and Novikov (MN; \cite{MN92}), the modified Kerr metric of Johannsen and Psaltis (MK; \cite{MKmetric}) and the modified-gravity, bumpy Kerr metric of Vigeland, Yunes, and Stein (MGBK; \cite{Vigeland:2011ji}). I aim to identify the manner in which their properties deviate from the special properties of the Kerr metric as a consequence of the no-hair theorem.

First I point out that the QK, BK, and MGBK metrics have been constructed as linear deviations from the Kerr metric \cite{GB06,CH04,VH10,Vigeland:2011ji}, while the MN and MK metrics are nonlinear deviations from the Kerr metric \cite{MN92,MKmetric}. The MN metric is Ricci flat, the QK metric is Ricci flat up to terms containing the quadrupole moment, and the BK metric is a vacuum solution of the linearized Einstein equations if the spin vanishes. The MK and MGBK metrics are not Ricci flat. On the other hand, the QK, BK, MN, and MK metrics are stationary and axisymmetric and are generally of Petrov type I, while the MGBK metric also admits an approximate Carter constant and is of approximate Petrov type D. All of these metrics are asymptotically flat, which I show explicitly in the case of the QK and MN metrics (see Refs.~\cite{Gair:2011ym,MKmetric} for the BK, MK, and MGBK metrics). I show, however, that the MN metric requires a coordinate transformation and a rescaling of the mass in order to reduce to the Newtonian limit in the nonrelativistic regime.

I proceed to investigate the nature of the central object in all five metrics. Using techniques from the numerical relativity literature, I calculate the location of event horizons. The horizon in each case can be expressed as a level surface of a particular scalar function (see Ref.~\cite{Thornburg}). For the cases I consider here, this function is governed by a differential equation of the horizon radius as a function of the polar angle, which I solve using both analytical and numerical methods. I also derive an approximate analytic expression of this equation for small perturbations away from the Kerr metric. I show that the QK and BK metrics harbor naked singularities as is the case of the MN metric \cite{MN92}, while the MGBK metric describes a black hole \cite{Gair:2011ym}. I also show that the MK metric harbors a naked singularity, which is located at the Killing horizon and can have either spherical or disjoint topology. I calculate expressions for the deviation parameter as a function of the spin that delineate the boundaries between the regions of the parameter space with these different topologies.

Finally, I identify the regions of space where violations of Lorentz symmetry or closed timelike curves exist, which I find outside of the central objects of the QK, BK, and MN metrics. These regions are unphysical and have to be excised by introducing a cutoff radius, which can, therefore, limit the ability of these metrics to serve as a framework for observational tests of the no-hair theorem. They impact both EMRI observations in the gravitational-wave spectrum, as well as electromagnetic observations of accretion flows, since both depend sensitively on the behavior of the metric near the innermost stable circular orbit (ISCO); see the discussion in Ref.~\cite{MKmetric}. Consequently, the QK, BK, and MN metrics can only be used for tests in the electromagnetic spectrum for moderately-spinning black holes so that the ISCO lies outside of the pathological regions.

I show that the MK and MGBK metrics are free of such pathologies exterior to the naked singularity and the event horizon, respectively, and argue that these metrics are particularly suited for electromagnetic and gravitational-wave tests, respectively. In the case of the MK metric, a cutoff radius has to be introduced just outside of the naked singularity. Since the ISCO lies outside of the naked singularity for all values of the spin and the deviation parameter \cite{MKmetric}, the cutoff radius can always be chosen so that the ISCO still lies in the domain exterior to the cutoff. The existence of a Carter-like constant in the MGBK metric allows one to separate the geodesic equations, which facilitates the study of EMRIs in such spacetimes. The MGBK is, thus, a useful tool for developing gravitational-wave models that can be used for tests of the no-hair theorem \cite{kludge,Vigeland:2011ji,Gair:2011ym}.

This paper is organized as follows: In Sec.~II, I compile the explicit forms of the five metrics that I study in this paper. In Sec.~III, I discuss their symmetries and show that they are asymptotically flat. I analyze in detail the presence of event horizons in the five metrics in Sec.~IV and identify pathological regions in Sec.~V. I formulate my conclusions and discuss astrophysical applications in Sec.~VI. Throughout, I use geometric units, where $G=c=1$.

\section{Parametric Deviations from the Kerr Metric}
\label{sec:pdKM}

In this section, I summarize the explicit form of the five spacetimes \cite{GB06,VH10,MN92,MKmetric,Vigeland:2011ji} that I use in this paper to investigate parametric deviations from the Kerr metric. 

My starting point is the Kerr metric $g_{\rm \mu\nu}^{\rm K}$, which in Boyer-Lindquist coordinates takes the form (e.g.~\cite{Bardeen72})
\begin{align}
g_{tt}^{\Kerr}&=-\left(1-\frac{2Mr}{\Sigma}\right)\,,
\nonumber \\
g_{t\phi}^{\Kerr} &= -\frac{2Mar\sin^2\theta}{\Sigma}\,,
\nonumber \\
g_{rr}^{\Kerr} &= \frac{\Sigma}{\Delta}\,,
\qquad
g_{\theta \theta}^{\Kerr} = \Sigma\,,
\nonumber \\
g_{\phi \phi}^{\Kerr} &= \left(r^2+a^2+\frac{2Ma^2r\sin^2\theta}{\Sigma}\right)\sin^2\theta\,,
\label{kerr}
\end{align}
where
\ba
\Delta &\equiv& r^2-2Mr+a^2\,, \qquad
\Sigma \equiv r^2+a^2\cos^2 \theta\,.
\label{deltasigma}
\ea
The only parameters in the Kerr metric are the first two multipole moments, i.e., the mass monopole $M_0=M$ and the spin dipole $S_1=J=aM$ of the black hole. All multipole moments of higher order are related to mass and spin by Eq.~(\ref{kerrmult}).

Some of the metric deformations that I study in this paper have been designed as linear deviations of the Kerr metric, which are of the form
\be
g_{\mu \nu} = g_{\mu \nu}^{\Kerr} + \zeta \, h_{\mu \nu}\,,
\label{eq:gen-form}
\ee
where $\zeta$ is a small parameter that reminds us that the metric deformation $\zeta h_{\mu \nu}$ is supposed to be small relative to the Kerr metric $g_{\mu \nu}^{\Kerr}$. These include the QK \cite{GB06}, BK \cite{CH04,VH10}, and MGBK \cite{Vigeland:2011ji} metrics. Unless I state otherwise, I perform my analysis of these metrics to linear order in the parameter $\zeta$, i.e., I expand all quantities that derive from a metric of the form given by Eq.~\eqref{eq:gen-form} to $\mathcal{O}(\zeta)$. Other parametric deviations from the Kerr metric need not be linear and can be more general functions of the deviation parameters. These include the MN \cite{MN92} and the MK \cite{MKmetric} metrics. I will analyze the properties of these metrics without expanding in the parameter $\zeta$ unless I state it explicitly.

For the study of astrophysical phenomena in the metrics with linear deviations from the Kerr metric, however, it is sometimes useful to compute their properties to all orders in the parameter $\zeta$, i.e., without expanding the results of such computations to $\mathcal{O}(\zeta)$. While an expansion in the small parameter $\zeta$ can always be performed in analytic calculations, it is a lot more difficult and, in some cases, even impossible to enforce in other settings such as the ones involving magnetohydrodynamic simulations, which numerically solve the (nonlinear) geodesic equations. For this purpose, I also study the existence of event horizons and pathological regions in the QK, BK, and MGBK metrics to all orders in the parameter $\zeta$. Note, however, that in this paper I only consider small values of the parameter $\zeta$. My results for the QK, BK, and MGBK metrics without expanding in the parameter $\zeta$ may be altered if larger values of the parameter $\zeta$ are considered. Similarly, it is instructive to study the MN and MK metrics also in the limit of small deviations, expanding these metrics to first order in the deviation parameter and treating the resulting metrics as perturbative.

\subsection{The quasi-Kerr metric}

The QK metric \cite{GB06} derives from the Hartle-Thorne metric \cite{HT}, which was originally designed for slowly rotating neutron stars. The QK metric deviates from the Hartle-Thorne metric in that its quadrupole moment is corrected, i.e., it is not assumed to depend on mass and spin through Eq.~\eqref{kerrmult}. Glampedakis and Babak \cite{GB06} and Johannsen and Psaltis \cite{PaperI} calculated orbital frequencies in this spacetime, and Johannsen and Psaltis \cite{PaperI,PaperIII} analyzed the properties of this spacetime including the locations of the ISCO and the circular photon orbit, the gravitational lensing experienced by photons, as well as the dynamical frequencies of thin accretion disks around the central object.

The QK metric \cite{GB06} modifies the quadrupole moment of the Kerr metric by the amount
\begin{equation}
\delta Q_{\rm QK} = -\epsilon_\QK M^3,
\end{equation}
where the parameter $\epsilon_\QK$ measures deviations from the Kerr metric. The full quadrupole moment is then
\begin{equation}
Q_{\rm QK} = -M\left(a^2+\epsilon_\QK M^2\right).
\label{qradmoment}
\end{equation}

In Boyer-Lindquist-like coordinates, i.e., in spherical-like coordinates that reduce to Boyer-Lindquist coordinates in the Kerr limit, the QK metric $g_{\rm \mu\nu}^{\QK}$ is given by Eq.~\eqref{eq:gen-form} with $\zeta_{\rm QK} \equiv \epsilon_\QK$ and \cite{GB06}
\ba
h_{\QK}^{tt}&=&(1-2M/r)^{-1}\left[\left(1-3\cos^2\theta\right)\mathcal{F}_1(r)\right], \nonumber \\
h_{\QK}^{rr}&=&(1-2M/r)\left[\left(1-3\cos^2\theta\right)\mathcal{F}_1(r)\right], \nonumber \\
h_{\QK}^{\theta\theta}&=&-\frac{1}{r^2}\left[\left(1-3\cos^2\theta\right)\mathcal{F}_2(r)\right], \nonumber \\
h_{\QK}^{\phi\phi}&=&-\frac{1}{r^2\sin^2\theta}\left[\left(1-3\cos^2\theta\right)\mathcal{F}_2(r)\right],
\label{h}
\ea
and $h_{\QK}^{t\phi}=0$. The functions $\mathcal{F}_{1,2}(r)$ are given in Appendix~A of Ref.~\cite{GB06}. The QK metric is a solution of the vacuum Einstein equations for spins that satisfy $a/M \ll 1$, provided $\epsilon_{\QK} \neq 0$.

\subsection{The bumpy Kerr metric}

The BK metric \cite{CH04,VH10} modifies the Kerr spacetime through small perturbations due to external stresses. Collins and Hughes~\cite{CH04} defined mass perturbations by starting from the most general stationary and spherically-symmetric metric in Weyl form
\begin{equation}
ds_{\rm W}^2 = -e^{2\psi}dt^2 + e^{2\gamma-2\psi}(d\rho^2 + dz^2) + e^{-2\psi}\rho^2d\phi^2.
\end{equation}
They defined $\psi \equiv \psi_0 + \psi_1$ and $\gamma \equiv \gamma_0 + \gamma_1$, with $(\psi_0,\gamma_0)$ equal to the Schwarzschild values and $(\psi_{1},\gamma_{1})$ parametric deformations. Both $\psi_1$ and $\gamma_1$ are proportional to the small parameter $\zeta_{\rm BK}$. They then required this metric to satisfy the Einstein equations to linear order in $\zeta_{\BK}$, thus obtaining differential equations for the deformation functions. 

This was generalized to a spinning spacetime~\cite{VH10} by applying a Newman-Janis rotation~\cite{NewmanJanis}. In Boyer-Lindquist-like coordinates, the BK metric is given by Eq.~\eqref{eq:gen-form} with~\cite{VH10}
\allowdisplaybreaks[3]
\begin{align}
h_{tt}^{\BK} &= -2\left(1-\frac{2Mr}{\Sigma}\right)\psi_1, \qquad
h_{tr}^{\BK} = -\gamma_1 \frac{2a^2Mr\sin^2 \theta}{\Delta\Sigma}, 
\nonumber \\
h_{t\phi}^{\BK} &= (\gamma_1-2\psi_1) \frac{2aMr\sin^2 \theta}{\Sigma}, \qquad
h_{rr}^{\BK} = 2(\gamma_1-\psi_1) \frac{\Sigma}{\Delta}, 
\nonumber \\
h_{r\phi}^{\BK} &= \gamma_1 \left[ 1 + \frac{2Mr(r^2+a^2)}{\Delta \Sigma} \right] a\sin^2 \theta, \nonumber \\
h_{\theta\theta}^{\BK} &= 2(\gamma_1-\psi_1)\Sigma, \nonumber \\
h_{\phi\phi}^{\BK} &= \left[ (\gamma_1-\psi_1) \frac{8a^2M^2r^2\sin^2 \theta}{\Delta\Sigma(\Sigma-2Mr)} - 2\psi_1 \left(1-\frac{2Mr}{\Sigma}\right)^{-1} \right] 
\nonumber \\
&\times \Delta\sin^2 \theta.
\label{b}
\end{align}

The perturbations $\psi_1$ and $\gamma_1$ satisfy linearized Einstein equations, which can be solved through a multipolar decomposition. At lowest order $(\ell=2)$, these functions are given by~\cite{VH10}
\begin{align}
\psi_1^{\ell=2}(r,\theta) &= \frac{B_2 M^3}{4} \sqrt{ \frac{5}{\pi} } \frac{1}{d(r,\theta,a)^3} \left[ \frac{3L(r,\theta,a)^2 \cos^2 \theta}{d(r,\theta,a)^2} -1 \right],
\nonumber \\
\gamma_1^{\ell=2}(r,\theta) &= B_2 \sqrt{ \frac{5}{\pi} } \left[\frac{L(r,\theta,a)}{2} \right.
\nonumber \\
\times
&\left. \frac{c_{20}(r,a) + c_{22}(r,a) \cos^2\theta + c_{24}(r,a) \cos^4\theta}{d(r,\theta,a)^5} -1 \right],
\label{psigamma}
\end{align}
where
\ba
d(r,\theta,a) &=& \sqrt{ |r^2 -2Mr + (M^2+a^2)\cos^2 \theta|}, \nonumber \\
L(r,\theta,a) &=& \sqrt{ (r-M)^2 + a^2\cos^2 \theta }, \nonumber \\
c_{20}(r,a) &=& 2(r-M)^4 - 5M^2(r-M)^2 + 3M^4, \nonumber \\
c_{22}(r,a) &=& 5M^2(r-M)^2 - 3M^4 + a^2[4(r-M)^2-5M^2], \nonumber \\
c_{24}(r,a) &=& a^2(2a^2+5M^2).
\label{BK_dfunction}
\ea
The strength of the perturbation to the Kerr metric at this order is determined by the parameter $\zeta_{\rm BK} \equiv B_2$. Note the absolute value signs in the function $d(r,\theta,a)$, which are missing from the corresponding expressions in Ref.~\cite{VH10}, Eq.~(5.6). This function is the translation into Boyer-Lindquist coordinates of the Weyl-sector function $\cosh^2 u \cos^2 v + \sinh^2 u \sin^2 v$. It is positive definite in the Weyl sector, and should be positive definite in Boyer-Lindquist coordinates as well; $\psi_1$ can become imaginary otherwise.

Vigeland and Hughes \cite{VH10} analyzed orbits and orbital frequencies in this spacetime, and Vigeland~\cite{Vigeland:2010xe} showed that, at lowest order, the perturbation changes the mass quadrupole, which is given by~\cite{Vigeland:2010xe}
\begin{equation}
Q_{\BK} = -M \left( a^2 + \frac{1}{2}B_2M^2\sqrt{ \frac{5}{\pi} } \right),
\end{equation}
so that
\begin{equation}
\delta Q_{\BK} = -\frac{1}{2}B_2M^3\sqrt{ \frac{5}{\pi} }.
\end{equation}

This equation suggests a formal relationship between the deformation parameter $\epsilon_\QK$ of the QK metric and the parameter $B_2$ of the BK metric via the relation
\begin{equation}
\epsilon_{\QK} = \frac{1}{2}B_2\sqrt{ \frac{5}{\pi} }\approx0.63 B_2.
\end{equation}
This mapping is somewhat misleading, however, as one might be tempted to conclude that the QK and BK parameterizations are equivalent when in fact they are not. The respective perturbations of the Kerr metric in the QK and BK metrics are different due to the different functional forms of the corrections $h_{\mu\nu}$ and $b_{\mu\nu}$ given by Eqs.~(\ref{h}) and (\ref{b}), respectively. In addition, for nonzero deviations from the Kerr metric, the QK metric is a solution of the vacuum Einstein equations only up to quadratic order in spin, while the BK metric is a solution of the vacuum Einstein equations, linearized in $\zeta_{\BK}$, to all multipole orders $l$.

\subsection{The Ricci flat metric proposed by Manko and Novikov}

The MN metric~\cite{MN92} is a nonlinear superposition of the Kerr metric with a static vacuum field that generalizes the former to a Ricci flat spacetime with arbitrary mass and current multipole moments. In its original form~\cite{MN92}, this metric harbors a naked singularity~\cite{MN92}. Gair et al.~\cite{Gair08} and Brink~\cite{Brink08} analyzed properties of orbits in the MN spacetime and found regions near the central singularity where geodesic motion becomes ergodic. They identified domains containing closed timelike curves around the origin which violate causality. In addition, Gair et al.~\cite{Gair08} computed the location of the ISCO and the orbital frequencies. Berti et al.~\cite{Berti05} analyzed the MN metric in the context of rotating neutron stars, and Bambi and Barausse~\cite{BambiBarausse,BambiBarausse2} and Bambi and Lukes-Gerakopoulos \cite{BLG13} studied accretion disks and their thermal emission and potential gravitational-wave signatures in an MN background, respectively. Contopoulos, Harsoula, and Lukes-Gerakopoulos \cite{CHL12} and Lukes-Gerakopoulos and Contopoulos \cite{LGC13} investigated the stability of periodic orbits in the MN metric and a possible observational signature thereof. Ergodic orbits in general non-Kerr spacetimes were analyzed by Refs.~\cite{Apostolatos09,Gerakopoulos}. 

In this paper, I will use a subclass of the MN metrics that describes electrically neutral compact objects and that is given by the line element of~\cite{MN92} transformed to Boyer-Lindquist-like coordinates
\begin{align}
ds^2_{\MN} &= -f_{\MN} \; dt^2 + 2 f_{\MN} \, \omega \,dt \,d\phi 
\nonumber \\
&+ \frac{ e^{2\Gamma} }{ f_{\MN} } \frac{ (r-M)^2 - (M^2 - a^2)\cos^2\theta }{ \Delta } dr^2 \nonumber \\
&+ \frac{ e^{2\Gamma} }{ f_{\MN} } \left[ (r-M)^2 - (M^2 - a^2)\cos^2\theta \right] d\theta^2 
\nonumber \\
&+ f_{\MN}^{-1} \left( \Delta\sin^2\theta -f_{\MN}^2\omega^2 \right) d\phi^2,
\label{MN-sph-metric}
\end{align}
with $f_{\MN}=f_{\MN}(r,\theta)$, $\omega=\omega(r,\theta)$. The quantities $\Gamma$, $f_{\MN}$ and $\omega$ are given in Appendix~A.

Following Ref.~\cite{Gair08}, I define 
\begin{equation}
q_{\MN} \equiv -\frac{ M_2 - M_2^{\Kerr} }{ M^3 }\,,
\end{equation}
which is a dimensionless parameter that measures the MN deviation from the Kerr quadrupole moment $M_2^{\Kerr}$.
The first few nonvanishing multipole moments are given by~\cite{MN92} (see, however, Ref.~\cite{Gair08} and references therein)
\begin{align}
M_0 &=M, \qquad
S_1 = aM, \nonumber \\
M_2 &=  -M(a^2 + q_{\MN}M^2), \nonumber \\
S_3 &= -aM(a^2 + 2q_{\MN}M^2).
\label{MNmoments}
\end{align}

Expanding the MN metric to first order in the parameter $q_{\rm MN}$, I obtain the linearized MN metric and identify $\zeta_{\rm MN} \equiv q_{\rm MN}$. Due to the lengthy form of this metric, I do not write it here explicitly.

\subsection{The modified Kerr metric proposed by Johannsen and Psaltis}
 
The metric proposed by Johannsen and Psaltis \cite{MKmetric} contains a set of free parameters which introduce nonlinear deviations from the Kerr metric. Johannsen and Psaltis~\cite{MKmetric} introduced polynomial corrections to the $(t,t)$ and $(r,r)$ elements of the Schwarzschild metric and transformed this ansatz into a Kerr-like metric via the Newman-Janis algorithm~\cite{NewmanJanis}. This metric is not Ricci flat.

The properties of iron lines, quasiperiodic variability, continuum spectra, and x-ray polarization from accretion disks in the MK metric have been analyzed in Refs.~\cite{PaperIV,Bambi2012,Bambi&Krawc}. Moreover, the topology of this metric as well as its implications for the properties of black-hole jets have been studied in Refs.~\cite{MKmetric,topology} and \cite{jets}, respectively. Other properties of this metric were analyzed in Refs.~\cite{ChenJing,Konoplya}.

In Boyer-Lindquist-like coordinates, the metric is given by the expression
\ba
ds_{\rm MK}^2 = && -[1+h(r,\theta)] \left(1-\frac{2Mr}{\Sigma}\right)dt^2 \nonumber \\
&& -\frac{ 4aMr\sin^2\theta }{ \Sigma }[1+h(r,\theta)]dtd\phi \nonumber \\
&& + \frac{ \Sigma[1+h(r,\theta)] }{ \Delta + a^2\sin^2\theta h(r,\theta) }dr^2 + \Sigma d\theta^2 \nonumber \\
&& + \left[ \sin^2\theta \left( r^2 + a^2 + \frac{ 2a^2 Mr\sin^2\theta }{\Sigma} \right) \right. \nonumber \\
&& + \left. h(r,\theta) \frac{a^2(\Sigma + 2Mr)\sin^4\theta }{\Sigma} \right] d\phi^2,
\label{metricMK}
\ea
where
\be
h(r,\theta) \equiv \sum_{k=1}^\infty \left( \epsilon_{2k} + \epsilon_{2k+1}\frac{Mr}{\Sigma} \right) \left( \frac{M^2}{\Sigma} \right)^{k}
\label{h(r,theta)}
\ee
with free parameters $\epsilon_k$. I will use this metric with only one nonzero parameter, so that the function $h(r,\theta)$ reduces to
\be
h(r,\theta) = \epsilon_3 \frac{M^3 r}{\Sigma^2}.
\label{hchoice}
\ee

Linearizing the MK metric to first order in the parameter $\zeta_{\rm MK} \equiv \epsilon_3$, I obtain the correction to the Kerr metric given by the expressions
\ba
h_{tt}^\MK &=& -\frac{rM^3(\Sigma-2Mr)}{\Sigma^3}, \nn \\
h_{rr}^\MK &=& \frac{rM^3(\Sigma-2Mr)}{\Delta^2 \Sigma}, \nn \\
h_{\theta\theta}^\MK &=& 0, \nn \\
h_{\phi\phi}^\MK &=& \frac{ra^2M^3(\Sigma+2Mr)\sin^4\theta}{\Sigma^3}, \nn \\
h_{t\phi}^\MK &=& -\frac{2ar^2M^4\sin^2\theta}{\Sigma^3}.
\label{MKlinearized}
\ea

\subsection{The modified gravity bumpy Kerr metric}

The MGBK metric, proposed by Vigeland, Yunes, and Stein~\cite{Vigeland:2011ji}, deforms the Kerr metric through certain bump functions, such that the resulting metric possesses three constants of the motion. Such a metric is also not Ricci flat. Vigeland, Yunes, and Stein~\cite{Vigeland:2011ji} analyzed orbits in this spacetime and showed that specific choices of the bump functions reproduce all known modified-gravity black hole solutions known to date. Approximate EMRI waveforms in this metric were constructed in Ref.~\cite{Gair:2011ym}.

In this paper, I use the class of the MGBK metrics studied in Ref.~\cite{Gair:2011ym}, where certain simplifications are made to guarantee certain properties (see Sec.~IIB in Ref.~\cite{Gair:2011ym}). With this at hand, the nonvanishing components of the MGBK metric in Boyer-Lindquist-like coordinates are as in Eq.~\eqref{eq:gen-form} with 
\begin{widetext}
\ba
	h^{\MGBK}_{tt} &=& -\frac{a}{M} \frac{\bar{P}^{}_2}{\bar{P}^{}_1} h^{\MGBK}_{t\phi} - \frac{a}{2M} \frac{\Sigma^2 \Delta}{\bar{P}^{}_1} \frac{\partial h^{\MGBK}_{t\phi}}{\partial r}  + \frac{(r^2+a^2)\hat{\rho}^{2} \Delta}{\bar{P}^{}_1} \bar{\gamma}_{1} 
			 + \frac{2a^2r^2 \Delta \sin^2\theta}{\bar{P}^{}_1} \bar{\gamma}_1 - \frac{a}{M} \frac{\Delta\sin^2\theta}{\Sigma} \frac{\bar{P}^{}_3}{\bar{P}^{}_1} \bar{\gamma}_3 + \frac{2\Delta}{\Sigma} \frac{\bar{P}^{}_4}{\bar{P}^{}_1} \bar{\gamma}_4 \nonumber \\
			&& - \frac{a^2}{2M} \frac{\Sigma \Delta^2 \sin^2\theta}{\bar{P}^{}_1} \frac{d\bar{\gamma}_1}{dr} - \frac{a}{2M} \frac{\Delta^2 (\hat{\Sigma}+2a^2Mr\sin^2\theta) \sin^2\theta}{\bar{P}^{}_1} \frac{d \bar{\gamma}_3}{dr} - \frac{a^2}{2M} \frac{\Delta^2 (\Sigma-4Mr) \sin^2\theta}{\bar{P}^{}_1} \frac{d\bar{\gamma}_4}{dr} \,, \nonumber \\
		h^{\MGBK}_{rr} &=& - \frac{\Sigma \bar{\gamma}_{1}}{\Delta}  \,,
\nonumber \\
	h^{\MGBK}_{\phi\phi} &=& -\frac{(r^2+a^2)^2}{a^2} h^{\MGBK}_{tt} + \frac{\Delta}{a^2} \Sigma \bar{\gamma}_{1} - \frac{2(r^2+a^2)}{a} h^{\MGBK}_{t\phi}  - \frac{2\Delta^2 \sin^2\theta}{a} \bar{\gamma}_3 + \frac{2\Delta^2}{a^2} \bar{\gamma}_4 \,, \nonumber \\
	\frac{\partial^2 h^{\MGBK}_{t\phi}}{\partial r^2} &=&  \frac{2a^2\sin^2\theta}{\Sigma^2} \frac{\bar{P}^{}_6}{\bar{P}^{}_1} h^{\MGBK}_{t\phi} - \frac{2r}{\Sigma}\frac{\bar{P}^{}_7}{\bar{P}^{}_1} h^{\MGBK}_{t\phi} + \frac{4aMr\sin^2\theta}{\Sigma}\frac{\bar{P}^{}_{15}}{\bar{P}^{}_{16}} \bar{\gamma}_{1} - \frac{4aMr\sin^2\theta}{\Sigma^2}\frac{\bar{P}^{}_8}{\bar{P}^{}_1} \bar{\gamma}_1  + \frac{2\sin^2\theta}{\Sigma^2}\frac{\bar{P}^{}_{10}}{\bar{P}^{}_1} \bar{\gamma}_3 \nonumber \\
		&& - \frac{16aM \sin^2\theta}{\Sigma^2}\frac{\bar{P}^{}_{11}}{\bar{P}^{}_1} \bar{\gamma}_4 - \frac{2a}{\Sigma^2}\frac{\bar{P}^{}_{12}}{\bar{P}^{}_1} \frac{d\bar{\gamma}_1}{dr} - \frac{2\sin^2\theta}{\Sigma^2}\frac{\bar{P}^{}_{13}}{\bar{P}^{}_1} \frac{d\bar{\gamma}_3}{dr} - \frac{2a\sin^2\theta}{\Sigma^2}\frac{\bar{P}^{}_{14}}{\bar{P}^{}_1} \frac{d\bar{\gamma}_4}{dr} - \frac{a \Delta \sin^2\theta}{\Sigma} \frac{d^{2} \bar{\gamma}_1}{dr^{2}} \nonumber \\
		&& - \frac{\Delta \sin^2\theta}{\Sigma^2}(\hat{\Sigma}+2a^2Mr\sin^2\theta) \frac{d^{2} \bar{\gamma}_3}{dr^{2}} - \frac{a \Delta (\Sigma-4Mr) \sin^2\theta}{\Sigma^2} \frac{d^{2} \bar{\gamma}_4}{dr^{2}} \,,
		\label{Carter-conds-DK}
\ea
\end{widetext}
where  
\ba
\hat{\rho}^{2} &\equiv& r^{2} - a^{2} \cos^{2}{\theta}\,,
\qquad
\\ 
\hat{\Sigma} &\equiv& (r^{2} + a^{2})^{2} -  a^{2} \Delta \sin^{2}{\theta}\,. 
\ea
and $\bar{P}^{}_i$ are polynomials in $(r,\cos\theta)$, given in the Appendix of Ref.~\cite{Vigeland:2011ji} (I adopt here the deformed Kerr parameterization of Ref.~\cite{Vigeland:2011ji}). 

The bumpy functions $\bar{\gamma}_{i} = \bar{\gamma}_{i}(r)$ depend on radius, and I parameterize this dependence via~\cite{Gair:2011ym}
\begin{align}
\bar{\gamma}_{A} &= \sum_{n=0}^{\infty} \gamma_{A,n} \left(\frac{M}{r}\right)^{n}\,,
\qquad
\label{gammaA}
\bar{\gamma}_{3} = \frac{1}{r} \sum_{n=0}^{\infty} \gamma_{3,n} \left(\frac{M}{r}\right)^{n}\,,
\end{align}
where $A$ is $1$ or $4$ and $(\gamma_{1,n},\gamma_{3,n},\gamma_{4,n})$ are constants that control the magnitude of the deformations. Additional simplifications~\cite{Gair:2011ym} allow us to set $\gamma_{1,0}=\gamma_{1,1}=\gamma_{3,0}=\gamma_{4,0}=\gamma_{4,1}=\gamma_{3,2} = 0$. 

The metric components $h^{\MGBK}_{tt}$ and $h^{\MGBK}_{\phi \phi}$ are fully determined once I solve the differential equation in Eq.~(\ref{Carter-conds-DK}) for $h_{t \phi}$. Doing so in a far-field expansion, I find
\ba
h^{\MGBK}_{t \phi} = M \sum_{n=2}^{N} h_{t\phi,n}(\theta) \left( \frac{M^n}{\Sigma^{n/2}} \right)\,,
\label{MGBK_htphi_series}
\ea
where the coefficients $h_{t\phi,n}$ are given in Ref.~\cite{Gair:2011ym}. Notice that the expressions presented here defer slightly from that of Ref.~\cite{Gair:2011ym}, as I use here a dimensional Kerr spin parameter $a$ and define $\Sigma$ and $\rho$ differently than in Ref.~\cite{Gair:2011ym}. I likewise write the bump functions as $\bar{\gamma}_i$ instead of $\gamma_i$ and the polynomials as $\bar{P}^{}_i$ instead of $P^{}_i$ to avoid confusion with the bump functions in the BK metric and the Legendre polynomials that occur in the MN metric (see Appendix~A).

In this paper, I study the lowest-order perturbations and only allow the coefficients $\gamma_{1,2},~\gamma_{3,1},~\gamma_{3,3}$, and $\gamma_{4,2}$ to be nonzero. This choice corresponds to setting $N=2$ in Eq.~(\ref{MGBK_htphi_series}).

\section{Symmetries and Asymptotic Flatness}
\label{asy-flat}

By construction, all the metrics described in Sec.~\ref{sec:pdKM} admit two Killing vectors corresponding to stationarity and axisymmetry~\cite{MN92,CH04,GB06,VH10,MKmetric}. This implies that these spacetimes possess an associated conserved energy and conserved ($z$ component of the) angular momentum. The MGBK metric possesses in addition an approximately conserved third quantity (a Carter-like constant) associated with the existence of an approximate Killing tensor (approximate in the sense that it satisfies Killing's equation to linear order in the deformation parameters). Therefore, the MGBK metric is of approximate Petrov type D, while the other four metrics (QK, MN, BK, and MK) are, in general, of Petrov type I.  

Some metrics considered here satisfy the vacuum Einstein equations, while others do not. The MN metric is Ricci flat, and thus, it is a solution of the vacuum Einstein equations. The QK metric is Ricci flat only up to second order in spin and first order in the perturbation parameter, i.e., neglecting terms of ${\cal{O}}(\epsilon_{\QK} a)$, ${\cal{O}}(a^{2})$ and ${\cal{O}}(\epsilon_{\QK}^2$). The BK metric is a vacuum solution only of the linearized Einstein equations in the limit $a\rightarrow0$, i.e., the Ricci tensor contains terms of ${\cal{O}}(a B_2)$ \cite{VH10}. The MK and MGBK metrics do not satisfy the vacuum Einstein equations and are not Ricci flat.

In order to make meaningful predictions of observables, spacetimes of black holes in isolation must be asymptotically flat, i.e., there must exist a coordinate system $(x^{0},x^{1},x^{2},x^{3})$ such that all metric components in these coordinates behave as $g_{\mu \nu} = \eta_{\mu \nu} + {\cal{O}}(1/r)$ as $r \to \infty$ in either spatial or null directions, where $\eta_{\mu\nu}={\rm diag}(-1,1,1,1)$ is the Minkowski metric and where $r$ is the Euclidean norm of the spatial coordinates~\cite{Beig}. In terms of the line element in Boyer-Lindquist-like coordinates, the subleading terms must fall off as~(e.g., \cite{Heusler96}): 
\begin{align}
ds^2 &= -\left[ 1-\frac{2M}{r} + \mathcal{O}\left(r^{-2}\right) \right]dt^2 
\nonumber \\
&- \left[\frac{4Ma}{r}\sin^2\theta + \mathcal{O}\left(r^{-2}\right) \right]dtd\phi + \bigg[1 + \mathcal{O}\left(r^{-1}\right) \bigg]
\nonumber \\
&\times \bigg[dr^2 + r^2\left(d\theta^2 + \sin^2\theta d\phi^2\right) \bigg].
\label{asympt}
\end{align}
Asymptotically flat spacetimes with a slower falloff cannot be stationary in general relativity~\cite{KM95}. The above definition of asymptotic flatness is not precise, due to issues with coordinate invariance and the precise way in which the $r \to \infty$ limit is taken (see Ref.~\cite{Wald:1984rg} for further details). However, this definition will suffice for my purposes in this paper.

For all of the metrics described in Sec.~\ref{sec:pdKM}, the Kerr part is clearly asymptotically flat; I am thus left with the task of showing that the deviations of these metrics from the Kerr metric do not spoil the asymptotic flatness of the background. The asymptotic flatness of the BK, MK, and MGBK metrics has already been shown \cite{MKmetric,Gair:2011ym}. I now turn to the QK and MN metrics.

\subsubsection{QK metric}

As an example of the QK metric, I consider the $(t,t)$ component, which has the form
\begin{align}
h_{tt}^{\QK} &\propto \frac{2M}{r^{2}} (2M^3 + 4M^2r - 9Mr^2 + 3r^3) 
\nonumber \\
&- 3 (r-2M)^2\ln\left(\frac{r}{r-2M}\right).
\end{align}
One can expand the logarithm in $r \gg M$ to show that $h_{tt}^{\QK} \propto -(16/5) M^{5}/r^{3}$, which clearly remains asymptotically flat. A similar argument holds for the other components of the QK metric.

\subsubsection{MN metric}

The MN metric should be considered separately, as here its asymptotic flatness is not as obvious. To see its structure more clearly, I perform the coordinate transformation
\be
r' \equiv  \exp \left[ \frac{ 4q_\MN }{ ( 1-\chi^2 )^{3/2} } \right] r, \quad
\phi' \equiv \exp \left[ - \frac{ 4q_\MN }{ ( 1-\chi^2 )^{3/2} } \right] \phi, 
\label{mntrafo}
\ee
with $(t,\theta)$ unchanged and $\chi \equiv a/M$. This transformation ensures that the MN metric reduces to the Minkowski spacetime at radial infinity. Transforming the metric, expanding its elements in $M/r \ll 1$ and linearizing them in $q_\MN \ll 1$, the metric perturbations become (dropping primes)
\ba
h_{tt}^\MN &=&  -\frac{8q_\MN M}{\left( 1-\chi^2 \right)^{3/2} } \frac{1}{r} + \mathcal{O}\left(r^{-2}\right),  
\nonumber \\
h_{t\phi}^\MN &=& \frac{32 \chi q_\MN M^2 \sin^2\theta}{\left( 1-\chi^2 \right)^{3/2} } \frac{1}{r} + \mathcal{O}\left(r^{-2}\right), 
\nonumber \\
h_{rr}^\MN &=& - \frac{8q_\MN M}{ \left( 1-\chi^2 \right)^{3/2} } \frac{1}{r}  \nn \\
&& - \frac{8 q_\MN M^2 (4-\chi^2 \sin^2 \theta)}{\left( 1-\chi^2 \right)^{3/2} } \frac{1}{r^{2}}+ \mathcal{O}\left(r^{-3}\right), 
\nonumber \\
h_{\theta \theta}^\MN &=& - \frac{8 q_\MN \chi^2 M^2 \cos^2 \theta}{ \left( 1-\chi^2 \right)^{3/2} } + \mathcal{O}\left(r^{-1}\right), 
\nonumber \\
h_{\phi \phi}^\MN &=& - \frac{8 q_\MN \chi^2 M^2 \sin^2 \theta}{\left( 1-\chi^2 \right)^{3/2} } 
+ \mathcal{O}\left(r^{-3}\right)\,.
\label{mnasympt2}
\ea
The parameter $M$ has to be rescaled in order for the MN metric to describe the correct Newtonian limit in the nonrelativistic regime:
\ba
M' &\equiv& M \exp \left[ - \frac{4q_\MN}{(1-\chi^2)^{3/2}} \right].
\label{MNmassrescaled}
\ea
This then eliminates all components of the metric perturbation to relative ${\cal{O}}(r^{-3})$, with the deformed metric becoming equal to the Kerr one with mass $M'$ and spin $a' = \chi M'$, and shows that the MN metric is indeed asymptotically flat. An investigation of a rescaling of the full metric and its parameters at ${\cal{O}}(M^{3}/r^{3},q_\MN^2)$ is beyond the scope of my analysis. From here on, I will use the MN metric in the form given by Eq.~(\ref{MN-sph-metric}) after applying the coordinate transformation in Eqs.~(\ref{mntrafo}) and the rescaling in Eq.~(\ref{MNmassrescaled}).

\section{The Event Horizon}

The event horizon of a black hole delineates the region of spacetime which cannot communicate with distant
observers (its interior) from the region which can communicate. In this section, I describe the calculation of the location of event horizons in stationary, axisymmetric, asymptotically flat metrics such as the ones listed in Sec.~II, using techniques developed in numerical relativity. I proceed to compute the location of the event horizons of these metrics should they exist.

The event horizon is a null surface, generated by null geodesics (``generators") that are trapped within that surface. The normal to a null surface, $n^\mu$, is itself null: $n^\mu n_\mu = 0$. I can take this surface to be the level surface of a
scalar function $f(x^\alpha)$, in which case the normal is simply $n^\mu = \nabla^\mu f = \partial^\mu f$. The event horizon is then defined by the
condition (see, e.g., Ref.~\cite{Thornburg} for a detailed discussion)
\be
g^{\mu\nu} (\partial_\mu f)(\partial_\nu f) = 0.
\label{masterhoreq}
\ee
Choosing a time coordinate, this becomes a quadratic equation for $\partial_t f$ which can be solved to show how the horizon evolves given an initial condition. This equation can also be evolved backwards given some final condition. Very powerful codes have been developed in the past decade which solve the ``master" horizon equation, Eq.~(\ref{masterhoreq}), in dynamical spacetimes. As long as the solution settles down to the Kerr metric at late times, these codes can find the level surface $f$ that describes the horizon quite robustly~\cite{Thornburg}.

One can distinguish between coordinate singularities, coordinate locations where the line element diverges, and true singularities by evaluating curvature scalars. I will here distinguish between these two cases by computing the Kretschmann scalar,
\be
K \equiv R_{\alpha\beta\gamma\delta} R^{\alpha\beta\gamma\delta},
\label{Kretschmann}
\ee
where $R^{\alpha}_{~\beta\gamma\delta}$ is the Riemann tensor.  

The metrics described in Sec.~II are all parameterized by spherical-like coordinates $(t, r, \theta, \phi)$ and are stationary and axisymmetric. In this case, the function $f$ which characterizes the horizon can then only depend on the coordinates $r$ and $\theta$, and I have
\be
g^{rr} (\partial_r f)^2 + 2g^{r\theta} (\partial_r f)(\partial_\theta f) + g^{\theta\theta}(\partial_\theta f)^2 = 0 .
\label{mastereqsimple}
\ee

Equation~(\ref{mastereqsimple}) defines event horizons in the spacetimes that I study. Note, however, that the existence of a solution of this equation is only necessary for the presence of an event horizon, but generally not sufficient, because solutions need not form a closed surface or can be singular, i.e., the Kretschmann scalar can diverge at this location. In these cases, a solution of Eq.~\eqref{mastereqsimple} is simply a null surface, but not an event horizon. In this paper, I will distinguish between the two as appropriate.

In the following, I will choose special forms for the level surface function $f$, which further simplify my analysis. I motivate this special form by first examining the Kerr spacetime, and then generalizing this to the spacetimes I consider here.

\subsection{Kerr black holes}

A useful first guess for the scalar function $f$ is the radial coordinate $r$. Since $f$ is only defined up to a constant
[only its derivatives enter Eq.~(\ref{masterhoreq})], it is useful to subtract a constant so that $f = 0$ on the horizon. Let us then set
\be
f = r - H ,
\ee
where $H$ is the (initially unknown) location of the spacetime's event horizon. Level surfaces of this function $f$ are a sequence
of nested coordinate spheres, and Eq.~(\ref{mastereqsimple}) simplifies to
\be
g^{rr}(H) = 0,
\label{grrzero}
\ee
which follows from choosing $f = 0$, or simply $r = H$, to define the horizon.

It is sometimes erroneously thought that this condition generically describes event horizons.
In fact, this is only true if surfaces of constant $r$ have uniform causal structure, i.e., if constant $r$ surfaces are everywhere
spacelike, null, or timelike (assuming that they are closed). If this is not the case, which depends
on the underlying coordinate system, then this condition will give the wrong solution. If constant
$r$ surfaces do have uniform causal structure, then I find that level surfaces of $f$ are spacelike at large radius, timelike
at small radius, and null at $r = H$. The classic black hole solutions of general relativity are of this type, at least in
the standard coordinates used to describe them. The solution in this case is given by
\be
H \equiv H_K \equiv r_+ = M + \sqrt{M^2 - a^2}
\label{kerrhorizon}
\ee
(ignoring the possibility of charge). Note that the Kerr horizon radius is typically denoted by $r_+$, while $H$ is often used in the numerical relativity literature to denote the horizon radius, which can vary with both time and position. I will use both notations in this section.

For any stationary and asymptotically flat spacetime, the event horizon is also a Killing horizon for some Killing vector $\chi^{\mu}$ if Hawking's rigidity theorem \cite{rigidity} or perhaps an appropriate generalization thereof (see, e.g., Ref.~\cite{rigidity1}) can be applied. Then, this Killing vector can be written as $\chi^{\mu} = t^{\mu} + \Omega \; \phi^{\mu}$ for some constant $\Omega$, where $t^{\mu}$ and $\phi^{\mu}$ are the temporal and azimuthal Killing vectors of the spacetime (see, e.g., Ref.~\cite{Carroll:2004st}). Requiring that this vector be null forces the condition
\be
g_{\phi \phi} \left(\Omega^{2} + 2 \Omega \frac{g_{t\phi}}{g_{\phi \phi}} + \frac{g_{tt}}{g_{\phi \phi}} \right) = 0\,,
\ee
which I can use to solve for the constant
\be
\Omega_{\pm} = - \frac{g_{t\phi}}{g_{\phi \phi}} \pm \sqrt{\frac{g_{t\phi}^{2}}{g_{\phi \phi}^{2}} - \frac{g_{tt}}{g_{\phi \phi}}}\,.
\ee
As one approaches the event horizon, the angular velocity $\Omega_{\pm}$ must approach a constant, $\Omega_{\pm}  \to - g_{t\phi}/g_{\phi \phi}$. This is because this constant represents the angular velocity of zero-angular momentum observers at the horizon, which must be single-valued. The only way this can happen is if (see, e.g., Ref.~\cite{Poisson:book})
\be
g_{t\phi}^2 - g_{tt} g_{\phi\phi} = 0\,.
\label{horizon1}
\ee
The radius at which Eq.~\eqref{horizon1} is satisfied defines the Killing horizon for the spacetime, since there $\chi^{\mu}$ is null. Equation~(\ref{horizon1}) is then equivalent to Eq.~(\ref{grrzero}). In general, however, the Killing and event horizons of a stationary spacetime are distinct (see, also, Ref.~\cite{Poisson:book}).

\subsection{Existence conditions for horizons of modified black holes}

In the spacetimes I consider, the metrics are more complicated than the Kerr metric, and the $f = r - H$ ansatz is not sufficient. I instead let the horizon radius be a function of $\theta$:
\be
f = r - H(\theta) .
\label{horizonf}
\ee
This form is sufficiently general to describe any horizon for which there is a unique horizon radius for any given angle
$\theta$; see Ref.~\cite{Thornburg} for discussion. This ansatz may not be adequate for extreme deformations or for horizons with
unusual topology (e.g., toroidal or disjoint horizons; see Fig.~5 of Ref.~\cite{Diener} for an example). It will, however, suffice for
my discussion in this paper as I will show in this section.

With this in mind, the horizon is then defined by the condition $r = H(\theta)$, where, using Eqs.~(\ref{mastereqsimple}) and (\ref{horizonf}), $H(\theta)$
is defined by the condition
\be
g^{rr} - 2g^{r\theta} \left( \frac{dH}{d\theta} \right) + g^{\theta\theta} \left( \frac{dH}{d\theta} \right)^2 = 0.
\label{hor_master}
\ee

The metric components which appear here are each functions of $r$ and $\theta$. I evaluate them at $r = H(\theta)$, and so Eq.~(\ref{hor_master}) is an ordinary differential equation for the horizon radius $H(\theta)$. A formulation of this kind is used in most numerical horizon
finders, modulo some small modifications to handle horizons of unusual topology (see Refs.~\cite{Thornburg,Diener} for detailed discussion). 

I next examine how to solve this equation for the particular cases that I study. In all of these cases, I work
in coordinates such that $g^{r\theta} = 0$, so I drop the cross term in Eq.~(\ref{hor_master}) in what follows.

\subsection{Linearly deviating spacetimes}

The QK, BK, and MGBK spacetimes are only specified as linear deviations from the Kerr metric; see Eq.~(\ref{eq:gen-form}). In order to analyze the existence of an event horizon in these metrics to linear order in the respective deviation parameters, I expand the function $H(\theta)$ as
\be
H(\theta) = H_K + \zeta \delta H(\theta),
\label{perturbedhor}
\ee
where I introduced the Kerr horizon radius $H_K$ here so I automatically find the Kerr solution for $\zeta=0$.
Using
\be
g^{\mu\nu} = g_{\Kerr}^{\mu\nu} - \zeta h^{\mu\nu},
\ee
Eq.~(\ref{hor_master}) becomes
\be
g_{\Kerr}^{rr} - \zeta h^{rr} + (g_{\Kerr}^{\theta\theta} - \zeta h^{\theta\theta}) \left( \zeta \frac{d\delta H}{d\theta} \right)^2 = 0.
\ee
Truncating at linear order and using $h^{rr} = g_{\Kerr}^{rr}g_{\Kerr}^{rr}h_{rr}$, I further simplify this equation to
\be
g_{\Kerr}^{rr} (1-\zeta g_{\Kerr}^{rr} h_{rr}) = 0
\label{horeqpert1}
\ee
which is equivalent to Eq.~(\ref{grrzero}) at $\mathcal{O}(\zeta)$. I now examine what this equation implies for the QK, BK, and MGBK metrics.

\subsubsection{QK metric}

In the QK metric, the element $g^{rr}_{\rm QK}$ is given by the expression
\ba
g^{rr}_{\rm QK} = && \frac{\Delta}{\Sigma} + \frac{5 \epsilon_\QK}{16r^2} (1-3\cos^2 \theta) \nn \\
&& \bigg[ 2\left( 3r^3 - 9M r^2 + 4M^2 r + 2M^3 \right) \nn \\
&& - 3r^2 \left( r^2 - 2M^2 \right) \ln \left( \frac{r}{r-2M} \right) \bigg].
\label{grrQK}
\ea
If $\epsilon_\QK\neq 0$, Eq.~(\ref{horeqpert1}) does not have a solution for all values of the angle $\theta$. Solutions only exist as long as $r>2M$; at radius $r=2M$ the logarithm in Eq.~(\ref{grrQK}) diverges. Evaluating the Kretschmann scalar $K$ given by Eq.~(\ref{Kretschmann}) and expanding $K$ to $\mathcal{O}(\epsilon_{\rm QK})$, I find that $K$ diverges at the radius $r=2M$, which I, thus, identify as a singularity. The solution of Eq.~(\ref{horeqpert1}), therefore, forms a null surface and not an event horizon, and the object is a naked singularity.

\begin{figure}[ht]
\begin{center}
\psfig{figure=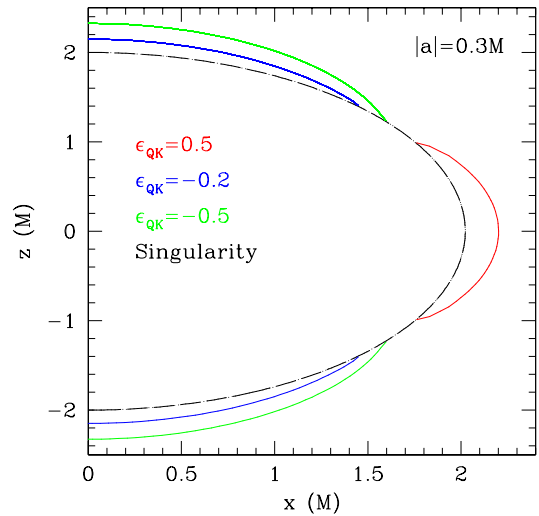,height=3.2in}
\end{center}
\caption{Null surface of the central object in the QK metric with a spin of $|a|=0.3M$ for several values of the parameter $\epsilon_\QK$. If $\epsilon_\QK\neq0$, the null surface is bound by the singularity located at radius $r=2M$, and the central object is a naked singularity.}
\label{horizonplot}
\end{figure}

For positive values of the parameter $\epsilon_\QK$, the null surface (where present) has a more oblate shape relative to the horizon of a Kerr black hole of equal spin, while the shape of the null surface is more prolate for negative values of the parameter $\epsilon_\QK$. The location of the null surface in the equatorial plane increases with increasing values of the parameter $\epsilon_\QK$ as already found in Ref.~\cite{PaperI}. There, however, the null surface was erroneously identified as the Killing horizon determined by the condition (\ref{horizon1}) leading to a slight difference in the location of this surface. In Fig.~\ref{horizonplot}, I plot, for illustrative purposes, the QK null surface for $|a|=0.3M$ and several values of the parameter $\epsilon_\QK$ in the $xz$ plane, where $x \equiv \sqrt{r^2+a^2}\sin\theta$ and $z \equiv r\cos\theta$.

In their construction of the QK metric, Glampedakis and Babak \cite{GB06} limited the validity of their metric to exclude the central region where $r\lesssim 2M$ due to the singularity located at radius $r=2M$. A corresponding cutoff radius $r_{\rm cutoff}(a,\epsilon_\QK) > 2M$ as a function of the spin and deviation parameter was defined heuristically in Ref.~\cite{PaperI} denoting an inner boundary of the region where the QK metric provides a consistent description of spacetime without pathologies.

\subsubsection{BK metric}

\begin{figure*}[htb]
\subfigure{\includegraphics[width=0.32\textwidth,clip=true]{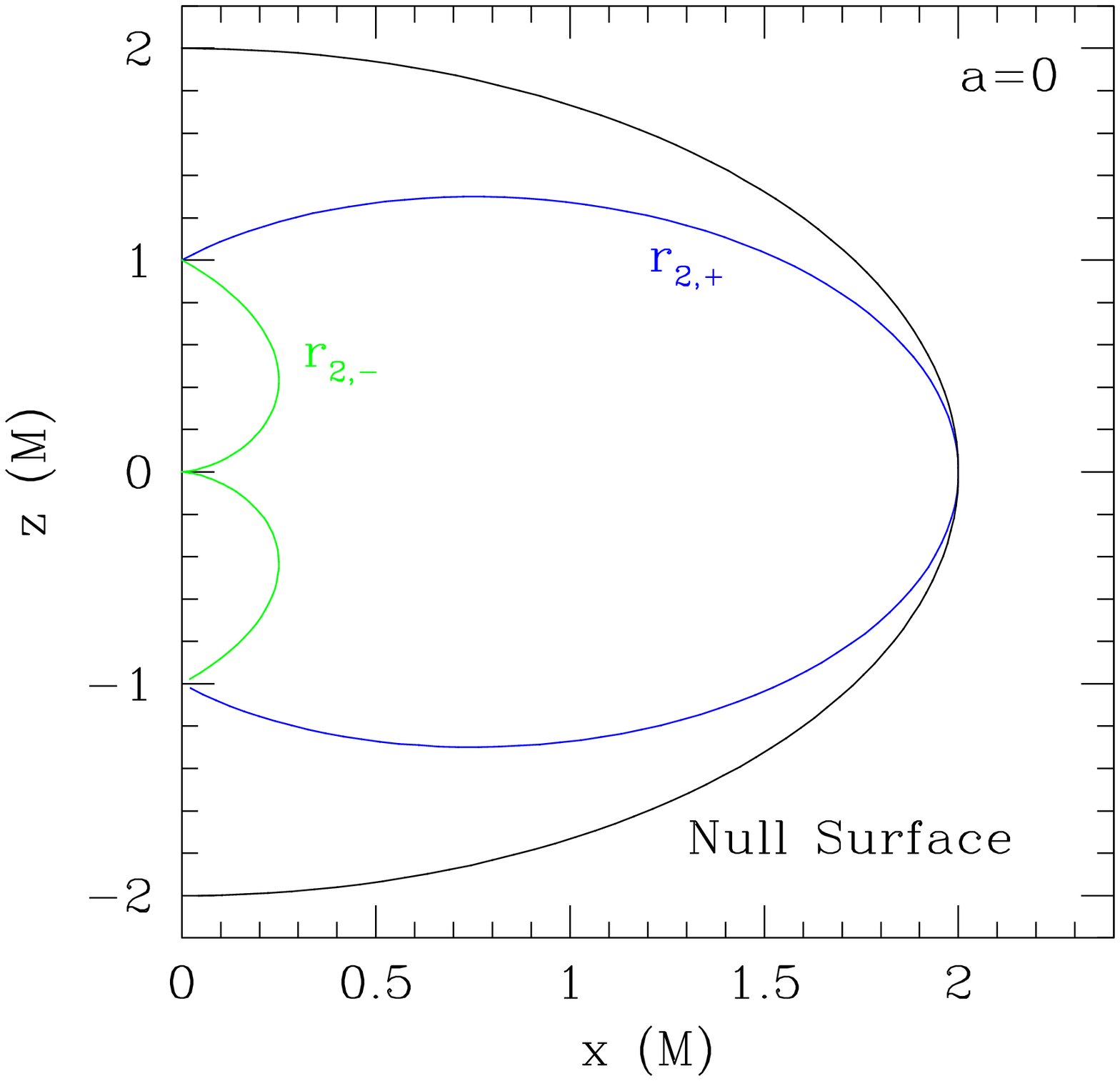}}
\subfigure{\includegraphics[width=0.32\textwidth,clip=true]{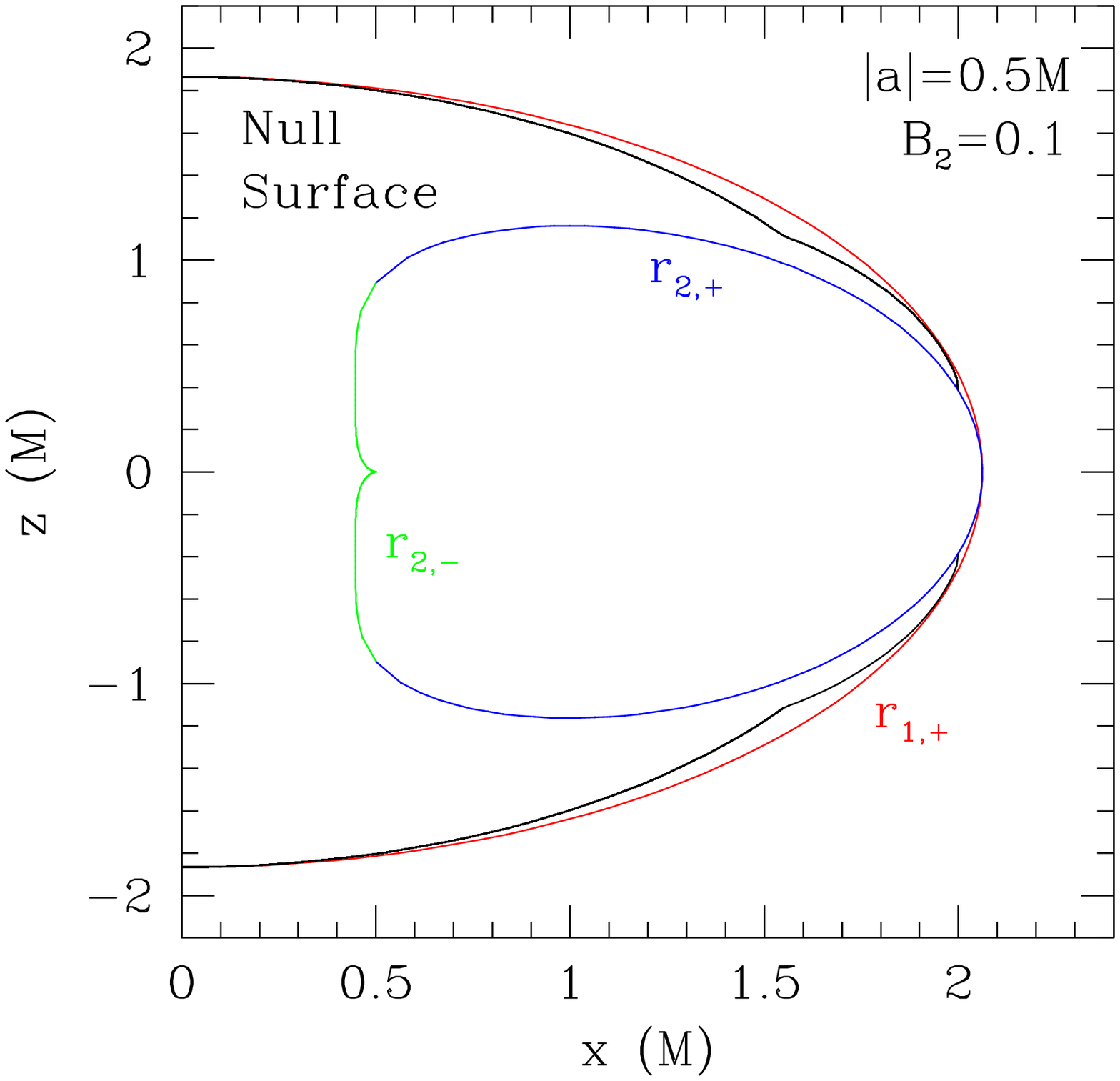}}
\subfigure{\includegraphics[width=0.32\textwidth,clip=true]{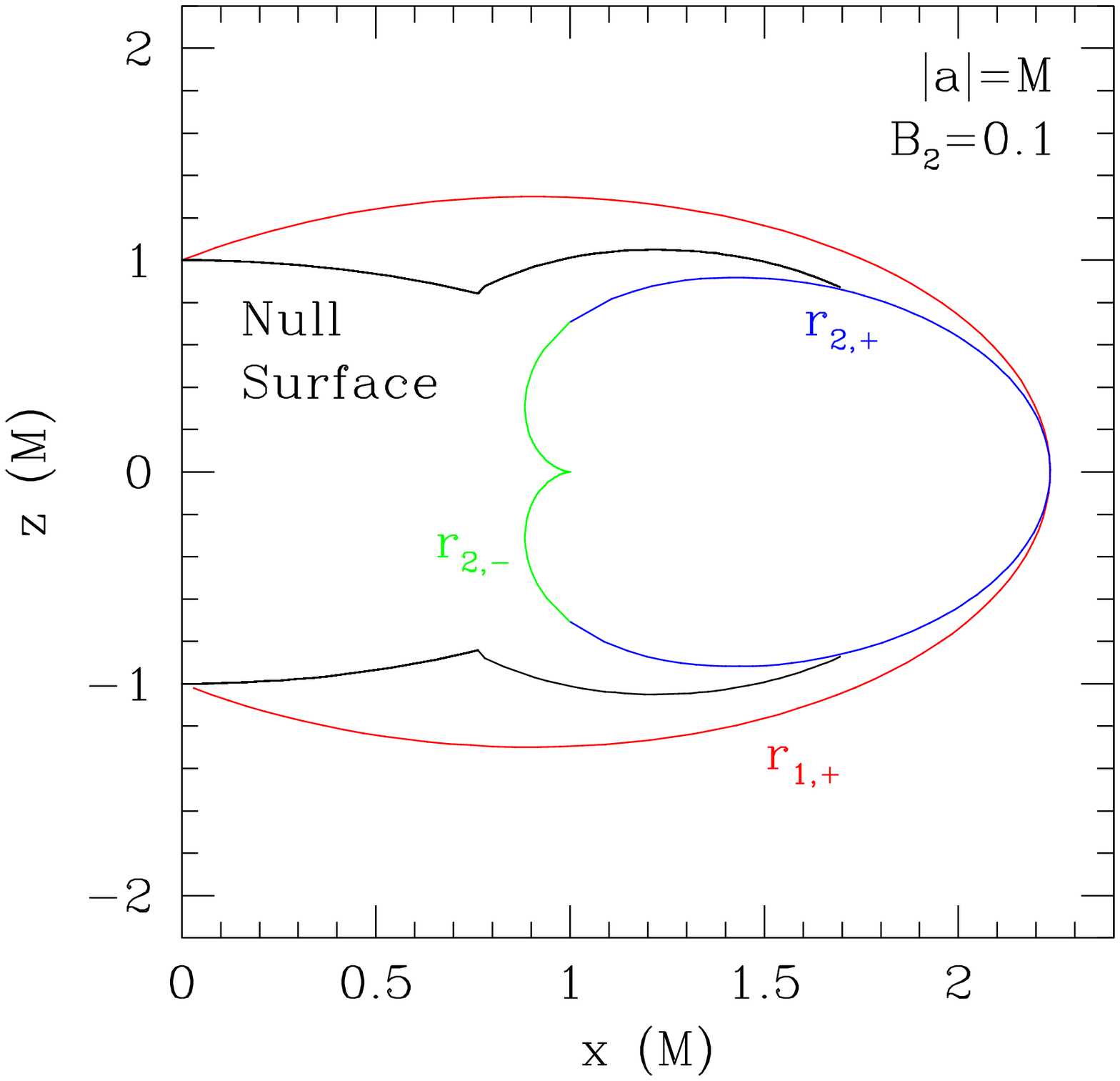}}
\caption{Location of the BK null surface and the singularities in the $xz$ plane for $a = 0$, $|a| = 0.5 M$, and $|a| = M$. For $a=0$, the singularity $r_{1,+}$ coincides with the null surface, and the singularities $r_{2,+}$ and $r_{2,-}$ are located inside the BK null surface. At $\theta=\pi/2$, the null surface intersects with the singularity $r_{2,+}$. As $|a|$ increases, the singularity $r_{1,+}$ moves outside of the BK null surface. For small but positive values of the spin $|a|$, the null surface is closed if $B_2>0$. Otherwise, it terminates at the singularity $r_{2,+}$. For all values of the deviation parameter $B_2\neq0$ the object constitutes a naked singularity.}
\label{rSing_BK_rth_plot}
\end{figure*}

If I look at the BK metric in Eq.~\eqref{b} and the definitions of the perturbation functions $\psi_1$ and $\gamma_1$ in Eq.~(\ref{psigamma}), I see that the metric becomes singular at $r_+$ given by Eq.~(\ref{kerrhorizon}), as well as where $\Sigma-2Mr=0$ and $d(r,\theta,a)=0$. The second condition occurs when $r$ is given by
\ba
r_{1,\pm} = M \pm \sqrt{M^2-a^2\cos^2\theta} \;,
\label{BK_sing1}
\ea
where I note that $r_{1,+} > r_{1,-}$, and $r_{1,+}$ coincides with the location of the Kerr ergosphere. The third condition occurs when $r$ is given by
\ba
r_{2,\pm} = M \pm \sqrt{M^2\sin^2\theta-a^2\cos^2\theta} \;,
\label{BK_sing2}
\ea
where clearly $r_{2,+} > r_{2,-}$. The Kretschmann scalar $K$ given by Eq.~(\ref{Kretschmann}) expanded to $\mathcal{O}(B_2)$ diverges at the singularities $r_{1,\pm}$ and $r_{2,\pm}$, which are, therefore, real singularities.

To leading order in the perturbation, the element $g^{rr}_{\rm BK}$ is
\be
g^{rr}_{\rm BK} = \frac{\Delta}{\Sigma} [ 1 - 2(\gamma_1 - \psi_1) ].
\ee
Consequently, as long as the bump functions $\gamma_{1}$ and $\psi_{1}$ are regular at $\Delta = 0$, then this last equality signals the location of a null surface. If, for some reason, the bump functions are singular here, then the null surface location would be modified from its Kerr value. Due to the presence of the singularities, the null surface cannot be an event horizon.

In Fig.\ \ref{rSing_BK_rth_plot} I plot the location of the null surface and the singularities in the $xz$ plane for $a = 0$, $|a| = 0.5 M$, and $|a| = M$. Since $a$ appears in the expressions for $r_{1,\pm}$ and $r_{2,\pm}$ as $a^2$, the locations of the singularities are the same for positive and negative values of the spin. For nonspinning black holes, the singularity $r_{1,+}$ coincides with the BK null surface. The singularity $r_{2,+}$ is hidden inside this surface at angles $0\leq \theta < \pi/2$ and coincides with this surface in the equatorial plane ($\theta=\pi/2$). As $|a|$ increases, the singularity $r_{1,+}$ moves outside of this surface. For small but nonzero values of the spin $|a|$, the null surface is closed if $B_2>0$ and terminates at the singularity $r_{2,+}$ near the equatorial plane if $B_2<0$ (see Fig.~\ref{BKvalidity} for an example). This equatorial ``hole" increases as $|a|$ increases.

The existence of naked singularities is consistent with the picture of the bumpy black holes originally described by Collins and Hughes \cite{CH04}, in which naked singularities of the Curzon type were explicitly introduced to change a spacetime's multipolar structure. Collins and Hughes~\cite{CH04} called attention to this strong-field naked singularity, cautioning that one could find odd results by using their spacetime to study very strong field structures. The same cautionary note clearly applies to the Vigeland and Hughes \cite{VH10} bumpy black hole as well. Though useful for studying  phenomena sensitive to orbits with separation $r > {\rm (a~few)} \times r_{1,+}$, the naked singularities that are introduced will surely have an adverse impact for orbits with $r \sim r_{1,+}$.

\subsubsection{MGBK metric}

For the MGBK metric, to leading order in the perturbation, the element $g^{rr}_{\rm MGBK}$ is
\be
g^{rr}_{\rm MGBK} = \frac{\Delta}{\Sigma} (1+\bar{\gamma}_1),
\ee
where $\bar{\gamma}_1$ is given by Eq.~(\ref{gammaA}). Similar to the case of the BK metric, the null surface is, then, determined by either the condition $\bar{\gamma}_1=-1$ or by the condition $\Delta=0$, where the latter condition holds if and only if $\bar{\gamma}_{1}$ remains regular at that radial location. In particular, if $\gamma_{1,n} \geq 0$ for $n \geq 2$, the null surface coincides with the Kerr event horizon $r_{\Hor}^{\MGBK} = r_+$. Moreover, all components of the metric and the inverse metric become singular only at $r=r_{+}$ or at $r=0$. However, by direct evaluation of the Kretschmann scalar, it is clear that $r=r_{+}$ is the location of a coordinate singularity, and not an essential one (the latter remains at the Kerr ring singularity). The null surface is, therefore, an event horizon. These results were also found in Sec.~II C of Ref.~\cite{Gair:2011ym}. In other limits of the MGBK metric (e.g., Refs.~\cite{Yunes:2011we,YYT12}), the location of the horizon is different from the location of the Kerr event horizon, as I demonstrate explicitly in Appendix~B.

\subsection{Spacetimes with nonlinear deviations}
\label{sec:exacthorizon}

The MN and MK spacetimes are nonlinear functions of the respective deviation parameters, and the full horizon equation \eqref{hor_master} has to be solved in order to study the existence of event horizons in these spacetimes. Likewise, one may be interested in the full solution of this equation to all orders in the parameter $\zeta$ for the QK, BK, and MGBK metrics. In this case, one no longer determines the location of a null surface or an event horizon using Eq.~\eqref{horeqpert1}, which is the version of Eq.~\eqref{hor_master} that has been linearized in the deviation parameter $\zeta$. In this section, I solve Eq.~(\ref{hor_master}) numerically for the horizon radius $H(\theta)$ for all five spacetimes. In Appendix~C, I describe two such methods, a spectral and a finite difference method, which I used for my analysis. First, however, I explore Eq.~(\ref{hor_master}) analytically taking advantage of the symmetries of these spacetimes.

Due to axisymmetry and reflection symmetry, the normal to the horizon must be purely radial at the poles ($\theta=0,\pi$) and in the equatorial plane ($\theta=\pi/2$). In these cases, the horizon equation, Eq.~(\ref{hor_master}), simplifies to Eq.~(\ref{grrzero}), which can be solved directly.

\subsubsection{QK metric}

For the QK metric, I find that the null surface coincides with the singularity at $r=2M$ at the poles and in the equatorial plane. Since the metric contains terms $\propto \ln[r/(r-2M)]$, which diverge at these locations, the numerical methods described in Appendix~C do not converge. However, I find another singularity located at the Killing horizon specified by Eq.~\eqref{horizon1} as long as it lies outside of the singularity at $r=2M$. For values of the parameter $\epsilon_\QK>0$, the Killing horizon lies around the equatorial plane, while for values of the parameter $\epsilon_\QK<0$, the Killing horizon lies around the poles.

\subsubsection{BK metric}

For the BK metric, I find that the null surface coincides with the singularity $r_{1,+}$ at the poles and that it does not pass through the equatorial plane if $B_2\neq0$. Due to the singularities, my numerical algorithms likewise did not converge, but I suspect that the null surface, where present, lies inside of the singularity $r_{1,+}$ as it does if the BK metric is treated perturbatively (at least for $|a|>0$).

\subsubsection{MGBK metric}

For the MGBK metric, the event horizon, again, coincides with the Kerr event horizon at the poles and in the equatorial plane. Numerically, I find that it is identical to the Kerr event horizon at all angles $\theta$ and that the Kretschmann scalar remains finite there.

\subsubsection{MN metric}

In prolate spheroidal coordinates $(t,x,y,\phi)$, the MN metric harbors a naked singularity \cite{MN92}. In Boyer-Lindquist-like coordinates, I find numerically that Eq.~(\ref{grrzero}) has no solution in the equatorial plane for any value of the deviation parameter $q_{\rm MN}\neq0$. I confirmed the absence of a horizon with the finite difference method, which did not converge if $q_{\rm MN}\neq0$.

However, by evaluating the metric elements, I find that the MN metric becomes singular at the location of the Kerr event horizon $r_+$, at which the elements $g_{rr}^{\rm MN}$ and $g_{\theta\theta}^{\rm MN}$ either diverge or vanish depending on the sign of the parameter $q_{\rm MN}$. Due to the lengthy form of the MN metric, I did not calculate its Kretschmann scalar to identify the nature of this singularity. For radii $r<r_+$, the MN metric can become imaginary (see Fig.~\ref{MNvalidity} for an example), and I, therefore, suspect the presence of a true singularity. My results remain if I linearize the MN metric to $\mathcal{O}(q_{\rm MN})$.

\subsubsection{MK metric}

The relevant components of the MK metric are given by the expressions
\ba
g^{rr}_{\rm MK} &=& \frac{ \Delta + a^2 \sin^2 \theta h(r,\theta) }{ \Sigma [ 1 + h(r,\theta) ] }, \\
g^{\theta\theta}_{\rm MK} &=& \frac{1}{\Sigma},
\ea
where the function $h(r,\theta)$ is given by Eq.~(\ref{hchoice}). First, I examine the metric element $g^{rr}_{\rm MK}$ at the poles and in the equatorial plane, again taking advantage of axisymmetry and reflection symmetry. If $\theta=0$ or $\theta=\pi$, $g^{rr}_{\rm MK}\propto \Delta$. This equation has a root at $r=H_K=r_+$ unless $\epsilon_3=-4r_+/M$, in which case the denominator likewise vanishes and $g^{rr}_{\rm MK}=1/2$. Therefore, the event horizon or null surface (if present) must pass through the $z$ axis at the Kerr horizon radius.

I explore next how any null surface must behave as I move in $\theta$ away from the $z$ axis. I examine $H(\theta)$ for small
angles, $\delta\theta \ll 1$, putting
\be
H(\delta\theta) = H_K + \delta\theta \frac{dH}{d\theta}\bigg|_{\theta=0} + \frac{\delta\theta^2}{2}\frac{d^2H}{d\theta^2}\bigg|_{\theta=0} .
\ee
Note that a similar expansion describes the behavior of $H$ near $\theta = \pi$; thanks to reflection symmetry, it is sufficient to focus on $\theta = 0$.

I insert $r = H(\delta\theta)$ into Eq.~(\ref{hor_master}), evaluate all metric functions at $\theta = \delta\theta$, and expand in $\delta\theta$. To ${\cal{O}}(\delta\theta)$, I find
\ba
0 = && \left( \frac{dH}{d\theta} \right) \bigg|_{\theta=0} \left[ \frac{(dH/d\theta)|_{\theta=0}}{2MH_K} + \frac{\delta\theta}{M} \left( \frac{ \sqrt{M^2-a^2} }{ H_K + M \epsilon_3/4 } \right.\right. \nn \\
&& + \left.\left. \frac{ (d^2H/d\theta^2)|_{\theta=0} }{ H_K } - \frac{ [(dH/d\theta)|_{\theta=0}]^2 }{ 2MH_K } \right) \right].
\ea
From this expression, I conclude that 
\be
\frac{dH}{d\theta} \bigg|_{\theta=0} = 0
\ee
in agreement with axisymmetry.

At second order in $\delta\theta$, I find a quadratic equation for $d^2H/d\theta^2$:
\ba
0 = && a^2 M \epsilon_3 + 4(M H_K - a^2) \frac{d^2H}{d\theta^2}\bigg|_{\theta=0} \nn \\
&& +(4H_K+M\epsilon_3) \left( \frac{d^2H}{d\theta^2} \bigg|_{\theta=0} \right)^2.
\ea
This equation only has real solutions as long as
\be
a^2 M \epsilon_3 (4H_K+M\epsilon_3) < 4(MH_K-a^2)^2 .
\label{eq:cond1}
\ee
Otherwise, $d^2H/d\theta^2$ is imaginary, and a null surface does not exist. Equation~\eqref{eq:cond1} implies an upper and a lower bound on the deviation parameter $\epsilon_3$, given by the expressions
\be
\epsilon_3^{\rm min-pole} < \epsilon_3 < \epsilon_3^{\rm max-pole},
\label{ep3boundpole}
\ee
where
\ba
\epsilon_3^{\rm max-pole} \equiv 2\frac{\sqrt{2MH_K-a^2}}{|a|} - \frac{2H_K}{M}, \label{ep3maxpole} \\
\epsilon_3^{\rm min-pole} \equiv -2\frac{\sqrt{2MH_K-a^2}}{|a|} - \frac{2H_K}{M}.
\label{ep3minpole}
\ea

In the equatorial plane, the condition given by Eq.~(\ref{grrzero}) reduces to
\be
r^2-2Mr+a^2+\epsilon_3 a^2 \frac{M^3}{r^3} = 0.
\label{MKgrreq}
\ee
There are no closed-form solutions to this quintic equation, but it is simple enough to find a numerical solution. Doing
so, I find generically that there is a maximum positive value of $\epsilon_3$ for which Eq.~(\ref{MKgrreq}) has a real solution.
To understand this $\epsilon_3$ threshold value, consider the shape of $g^{rr}_{\rm MK}(\theta = \pi/2)$ as a function of $r$. For modest positive
values of $\epsilon_3$, it has two roots which tend to be close to $r_\pm=M\pm\sqrt{M^2-a^2}$. There is also a minimum at
\be
r_{\rm min} = \frac{4M}{5} \left( 1 + \sqrt{ 1-\frac{15}{16} \frac{a^2}{M^2}  } \right),
\ee
which is roughly halfway between these roots. As $\epsilon_3$ is increased, the roots move toward the minimum, with all three
points converging at a value $\epsilon_3^{\rm max-eq}$ given by
\be
g^{rr}_{\rm MK}(r = r_{\rm min},\theta = \pi/2, \epsilon_3^{\rm max-eq} ) = 0.
\ee
Solving this equation yields
\ba
\epsilon_3^{\rm max-eq} = && \frac{1}{3125(a/M)^2} \bigg[ 1024 \left( 4 + \sqrt{16-15(a/M)^2} \right) \nn \\
&& - 160(a/M)^2 \left(40+7\sqrt{16-15(a/M)^2} \right) \nn \\
&& + 150 (a/M)^4 \left(15+\sqrt{16-15(a/M)^2} \right) \bigg].
\label{ep3maxeq}
\ea
Therefore, a null surface can only exist for values of the deviation parameter $\epsilon_3 < \epsilon_3^{\rm max-eq}$ given by Eq.~(\ref{ep3maxeq}). The bound $\epsilon_3^{\rm max-eq}$, however, is weaker than the condition given in Eq.~(\ref{ep3boundpole}), since $\epsilon_3^{\rm max-eq} > \epsilon_3^{\rm max-pole}$.

Imagine now that a null surface does in fact pass through the equator, and consider how it behaves as I move $\delta\theta$ away from the equatorial plane.
Let $H_{\rm eq}$ be the null surface radius at $\theta = \pi/2$ and put
\be
H(\pi/2 + \delta\theta) = H_{\rm eq} + \delta\theta \frac{dH}{d\theta}\bigg|_{\theta=\pi/2} + \frac{\delta\theta^2}{2} \frac{d^2H}{d\theta^2}\bigg|_{\theta=\pi/2}.
\ee
I insert $r = H(\pi/2+\delta\theta)$ into Eq.~(\ref{hor_master}) and put $\theta = \pi/2+\delta\theta$. Note that in this case, I cannot do as much analytic exploration. Because $H_{\rm eq}$ must itself be solved numerically, much of my analysis must likewise be numerical.

I find that $(dH/d\theta)|_{\theta=\pi/2} = 0$ as expected for all parameters examined and that a real solution for $(d^2H/d\theta^2)|_{\theta=\pi/2}$ exists for all values $0<\epsilon_3<\epsilon_3^{\rm max-pole}$. For $\epsilon_3<0$ and $|a|\gtrsim0.82M$, however, there exists a part of the parameter space in the range $\epsilon_3^{\rm min-pole}<\epsilon_3<0$ within which there is no real solution for $d^2H/d\theta^2$. Therefore, a null surface cannot exist in this region either.

In order to further analyze the nature of the central object in the MK metric, I calculate the location of the Killing horizon using Eq.~\eqref{horizon1}. Setting $\theta=0$ or $\theta=\pi$ in Eq.~\eqref{horizon1}, I find
\be
\left[ g_{tt}^{\rm MK}g_{\phi\phi}^{\rm MK} - \left(g_{t\phi}^{\rm MK}\right)^2\right] \propto \Delta \left[ 1 + \epsilon_3 \frac{rM^3}{\left(r^2+a^2\right)^2} \right].
\label{Killing_poles}
\ee
Therefore, the Killing horizon coincides with the Kerr event horizon (and hence with the null surface, if it exists) at the poles. For values of the parameter $\epsilon_3\leq\epsilon_3^{\rm Kil-pol}$, where
\be
\epsilon_3^{\rm Kil-pol} \equiv - \frac{16}{3\sqrt{3}} \left( \frac{a}{M} \right)^3,
\ee
Killing horizons in addition to the (outer and inner) Killing horizons emerge near the origin; one such horizon emerges if $\epsilon_3=\epsilon_3^{\rm Kil-pol}$ and two if $\epsilon_3<\epsilon_3^{\rm Kil-pol}$.

Equation~\eqref{horizon1} in the equatorial plane ($\theta=\pi/2$) reduces to
\be
\left( 1 + \epsilon_3 \frac{M^3}{r^3} \right) \left( \Delta + \epsilon_3 \frac{a^2M^3}{r^3} \right) = 0,
\label{Killing_plane}
\ee
where the second factor is identical to Eq.~\eqref{MKgrreq}, which determines the location of the null surface in the equatorial plane. This equation has two solutions if $\epsilon_3<\epsilon_3^{\rm max-eq}$ [c.f., Eq.~\eqref{ep3maxeq}], which are identical to the locations of the null surfaces (i.e., the location of the inner and outer null surface if $0\leq\epsilon_3<\epsilon_3^{\rm max-eq}$ and of the null surface if $\epsilon_3<0$, in which case there exists only one). If $\epsilon_3=-8$, the inner and outer Killing horizon coincide in the equatorial plane, and if $\epsilon_3<-8$, these two horizons cross, i.e., the inner Killing horizon lies outside of the outer Killing horizon in and around the equatorial plane, while it lies inside near the poles. For values of the parameter $\epsilon_3\geq\epsilon_3^{\rm max-eq}$, the Killing horizon does not pass through the equatorial plane, and the Killing horizon is disjoint forming two spherical surfaces centered on the symmetry axis above and below the equatorial plane.

Numerically, I find that a null surface exists if $0.1\gtrsim\epsilon_3\gtrsim-8$. In this region of the parameter space, the Killing horizon lies slightly outside of the null surface at the polar angles $0<\theta<\pi/2$ and $\pi/2<\theta<\pi$ if $\epsilon_3\neq0$. Both surfaces coincide at all polar angles only if $\epsilon_3=0$, in which case they form the Kerr event horizon as encapsulated in Hawking's rigidity theorem \cite{rigidity}. Outside of this region, the existence of a null surface is uncertain due to increasing numerical error in the solution of Eq.~\eqref{hor_master}.

Evaluating the Kretschmann scalar of the MK metric, I find that it diverges at the Killing horizon at all angles $0<\theta<\pi$. Due to the polar coordinate singularity of the Boyer-Lindquist-like coordinates, it is unclear if the Kretschmann scalar likewise diverges at the poles of the Killing horizon. The Killing horizon exists for all values of the spin and the parameter $\epsilon_3$ even if there is no null surface. Since the Killing horizon is singular if $\epsilon_3\neq0$, the MK metric harbors a naked singularity located at the Killing horizon.

\begin{figure}[ht]
\begin{center}
\psfig{figure=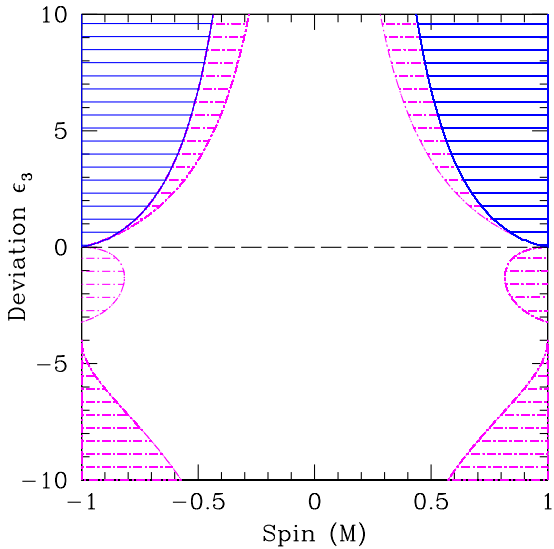,height=3.2in}
\end{center}
\caption{Region of the parameter space of the MK metric where a null surface exists. In this and the shaded regions, for values of the parameter $\epsilon_3\neq0$, the central object is a naked singularity located at the Killing horizon, which is of spherical topology. If a null surface exists, the Killing horizon coincides with the null surface at the poles and in the equatorial plane and lies outside of the null surface otherwise. In the blue shaded regions, the Killing horizon is of disjoint topology. Due to numerical uncertainties in the determination of the location of the null surface, I find that a null surface may not exist in this region if $\epsilon_3\lesssim-8$ or if $\epsilon_3\gtrsim0.1$. The black dashed line corresponds to a Kerr black hole.}
\label{MKallowedregion}
\end{figure}

\begin{figure*}[ht]
\begin{center}
\psfig{figure=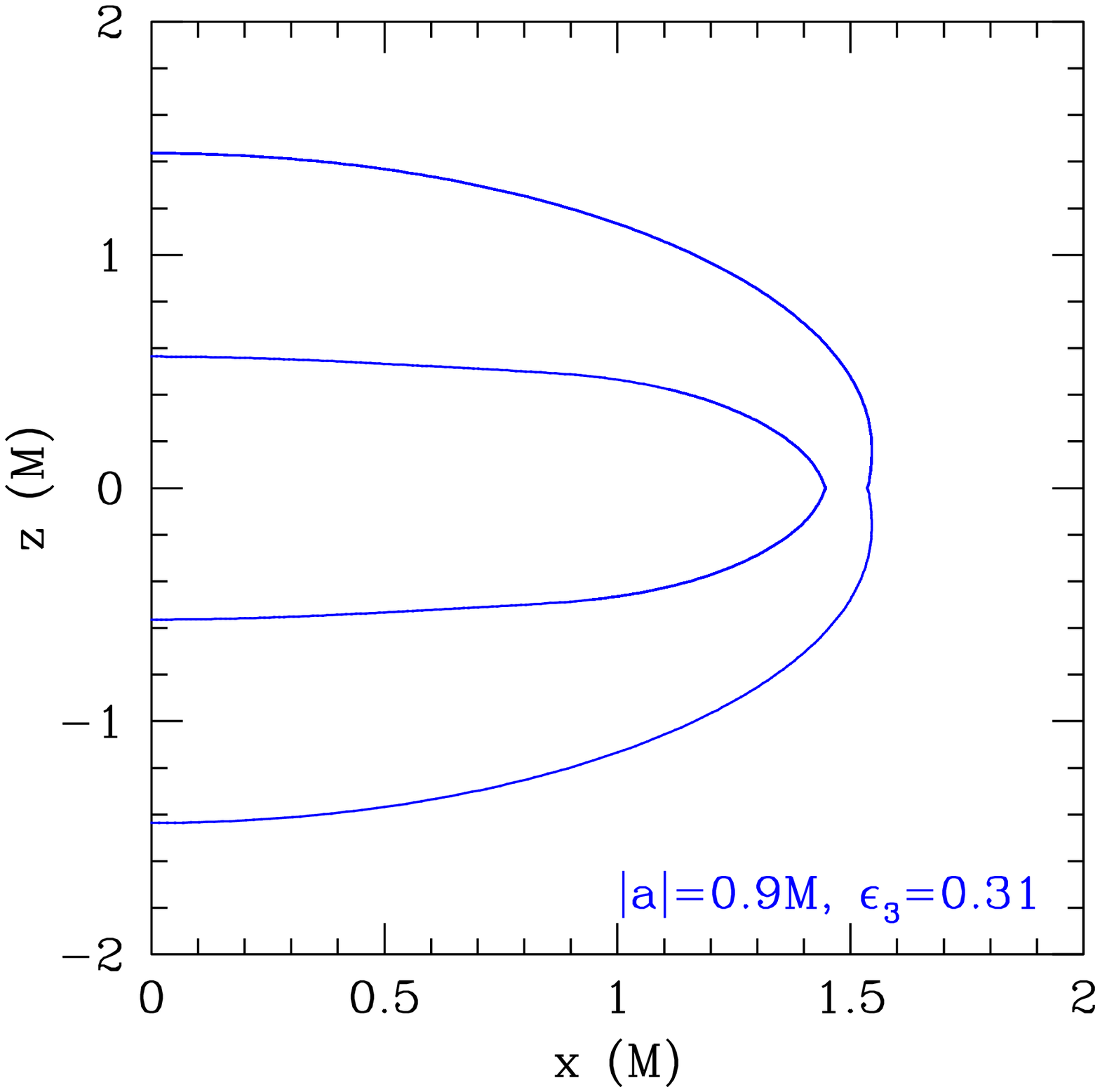,width=0.32\textwidth}
\psfig{figure=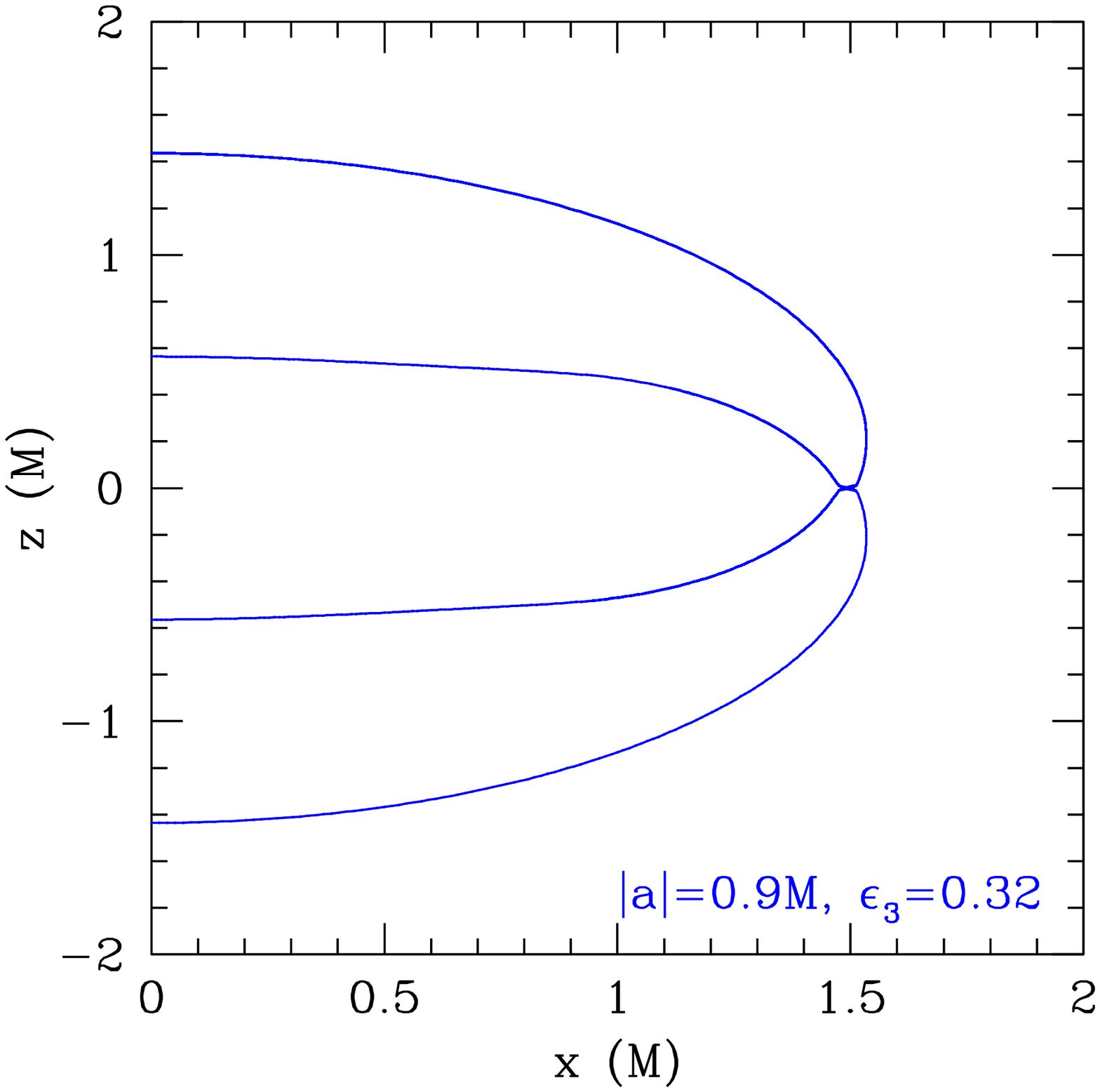,width=0.32\textwidth}
\psfig{figure=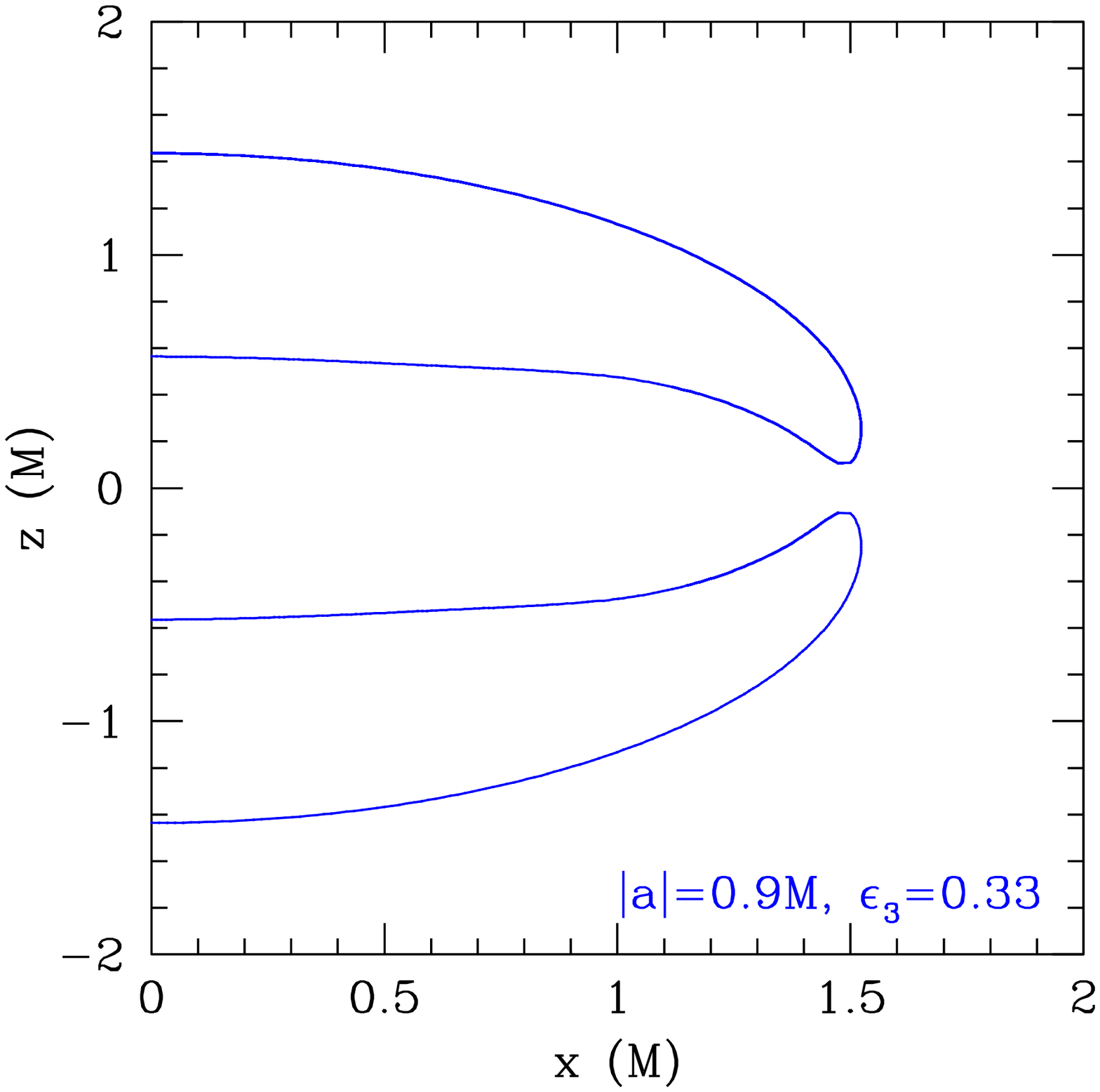,width=0.32\textwidth}
\end{center}
\caption{Different shapes of the Killing horizon in the MK metric for values of the spin $|a|=0.9M$ and deviation parameter $\epsilon_3$. Left panel: $\epsilon_3=0.31<\epsilon_3^{\rm max-eq}(a)$, where $\epsilon_3^{\rm max-eq}(a)$, $a>0$, denotes the boundary between the regions of the parameter space with different topologies of the Killing horizon. The Killing horizon has spherical topology consisting of an inner and outer sphere. The event horizon is located outside of the outer Killing horizon, and the central object is a black hole. Center panel: $\epsilon_3=0.32\approx\epsilon_3^{\rm max-eq}(a)$. The inner and outer Killing horizons merge in the equatorial plane, and the event horizon vanishes. Right panel: $\epsilon_3=0.33>\epsilon_3^{\rm max-eq}(a)$. The Killing horizons about the origin have split into two sphere-like surfaces located above and below the equatorial plane, respectively. In the central and right panels, the central object is a naked singularity located at the Killing horizon. For all values of the parameter $\epsilon_3$, the origin is likewise singular.}
\label{topochange09}
\end{figure*}

In Fig.~\ref{MKallowedregion}, I plot the various regions of the parameter space of the MK metric. At a given value of the spin, the shapes of the Killing horizon (if it has spherical topology) and, if present, the null surface are more prolate than the Kerr event horizon for values of the parameter $\epsilon_3>0$, while they are more oblate for values of the parameter $\epsilon_3<0$ (see Ref.~\cite{MKmetric}). I plot illustrative examples of the topology transition of the Killing horizon across the boundary $\epsilon_3^{\rm max-eq}(a)$ given by Eq.~\eqref{ep3maxeq} in Fig.~\ref{topochange09}.

The bound $\epsilon_3^{\rm max-eq}(a)$ coincides with the bound found in Ref.~\cite{MKmetric}, which mistakenly assumed the null surface to be a Killing horizon and used Eq.~(\ref{horizon1}) instead of Eq.~(\ref{hor_master}). As shown in Ref.~\cite{MKmetric}, for positive values of the spin this boundary also separates the regions of the parameter space where no ISCO exists [i.e., the blue region on the right-hand side in Fig.~\ref{MKallowedregion}, where $\epsilon_3\geq\epsilon_3^{\rm max-eq}(a)$, $a>0$] from the region where an ISCO exists (i.e., everywhere else in Fig.~\ref{MKallowedregion}). If an ISCO exists, it is always located outside of the Killing horizon and, therefore, outside of any region containing singularities \cite{MKmetric}. If an ISCO does not exist, the only singularity in the equatorial plane is the origin; I analyze the corresponding part of the parameter space in detail in a separate paper \cite{Joh13}. Reference~\cite{topology} used the condition given by Eq.~(\ref{grrzero}) to locate the event horizon, which likewise did not take the proper angular dependence of the horizon into account. 

\begin{figure*}[ht]
\begin{center}
\psfig{figure=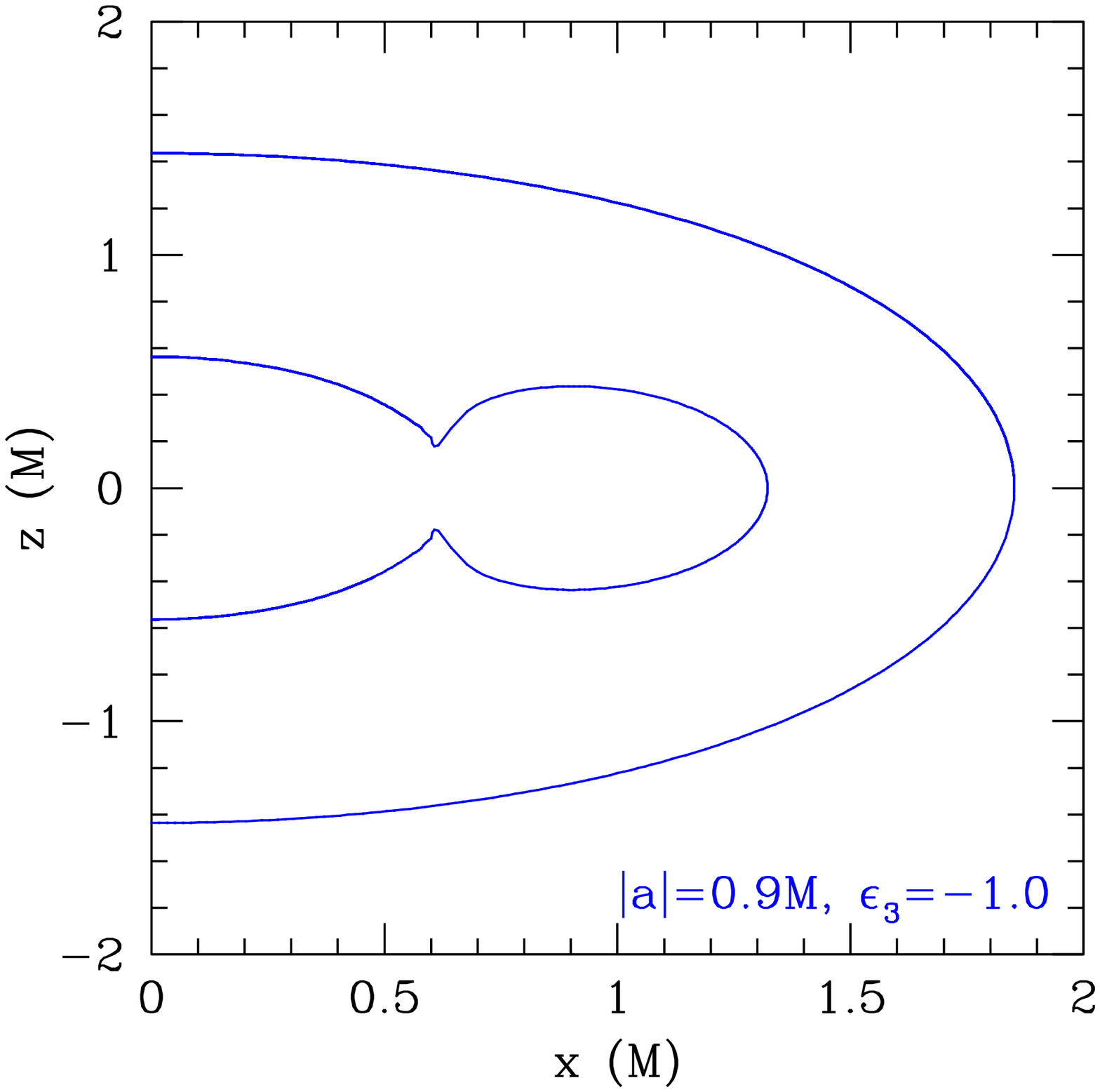,width=0.32\textwidth}
\psfig{figure=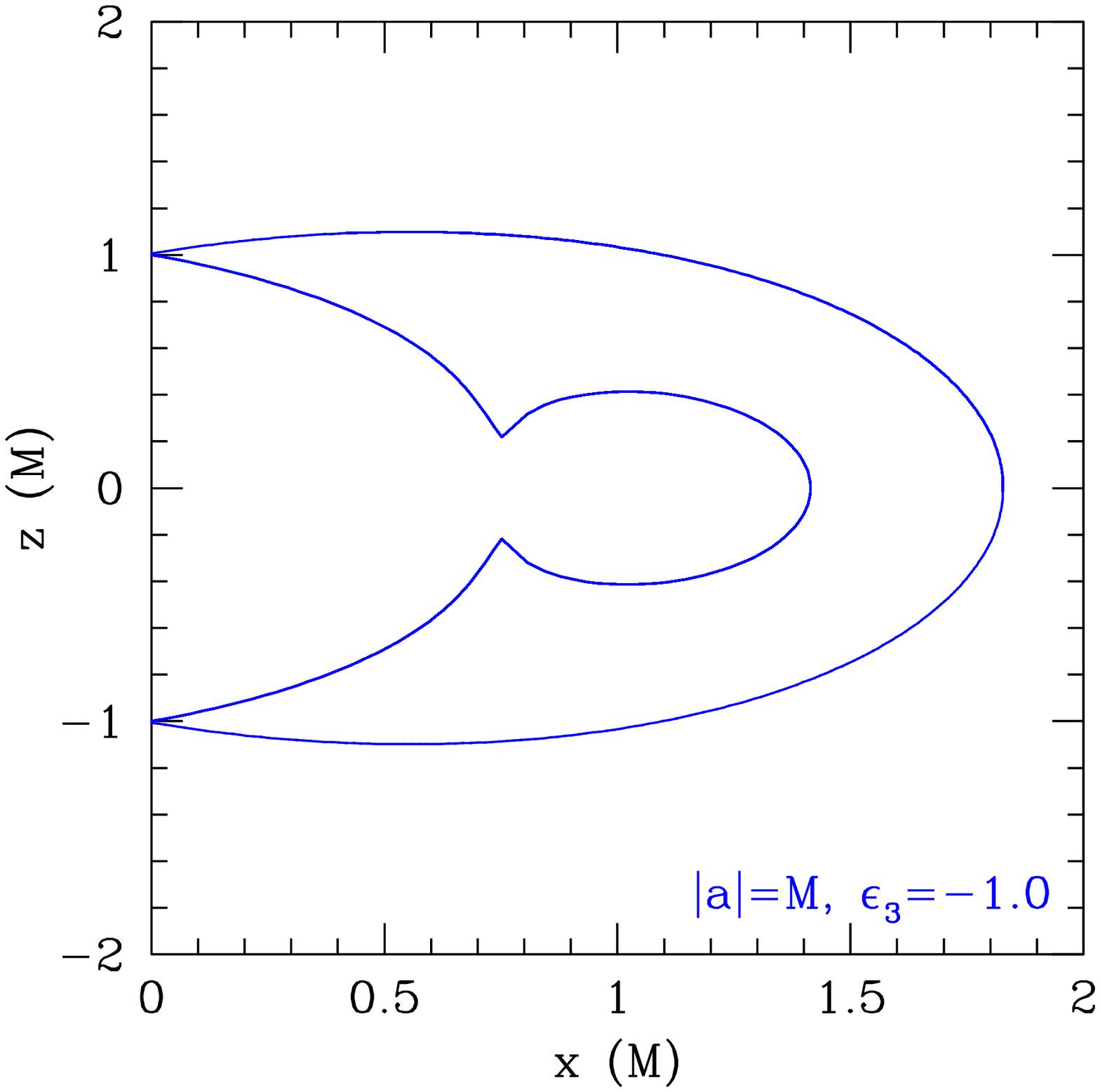,width=0.32\textwidth}
\psfig{figure=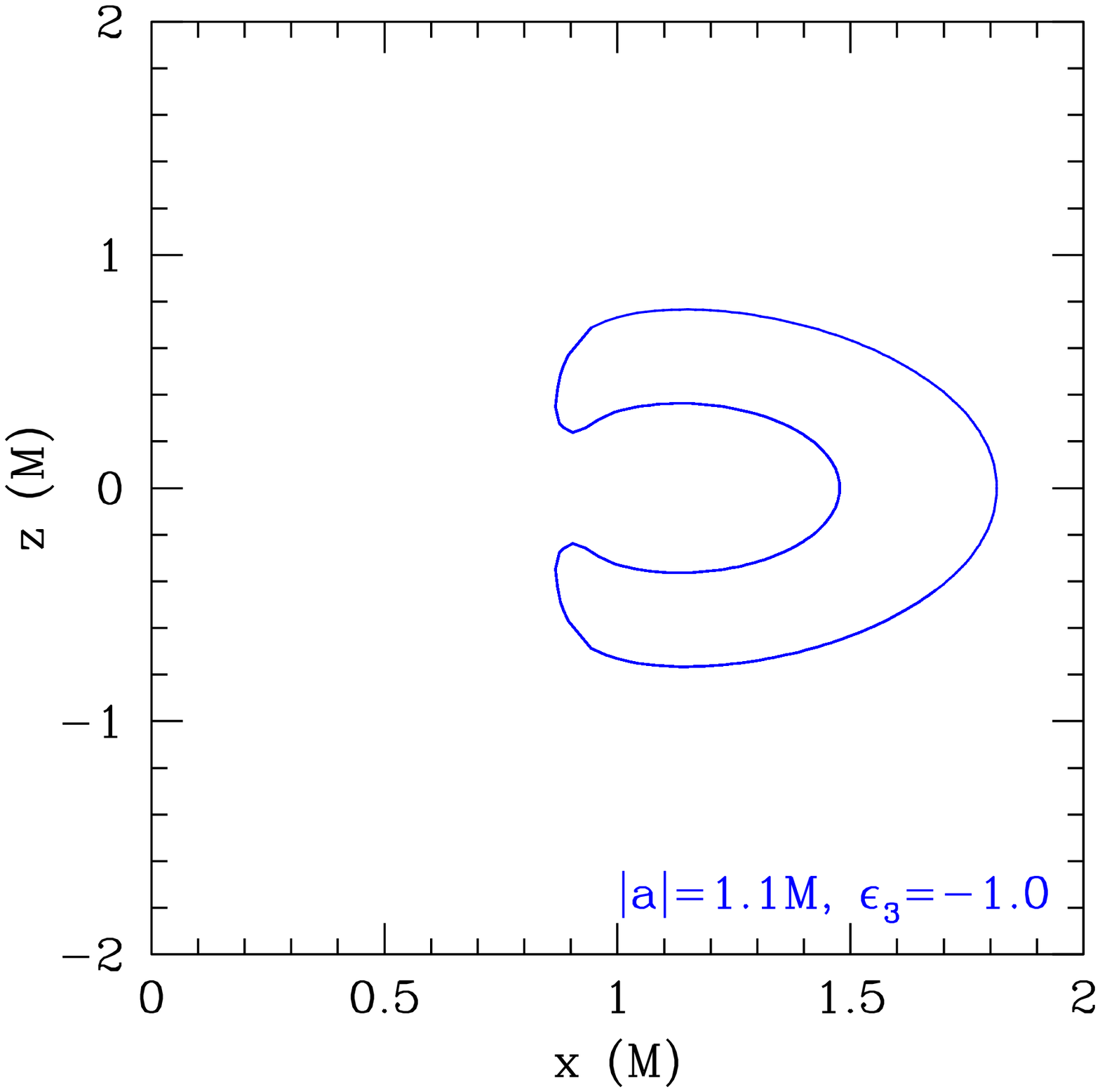,width=0.32\textwidth}
\end{center}
\caption{Different shapes of the Killing horizon in the MK metric in the case $\epsilon_3=-1$ for several values of the spin. Left panel: $|a|=0.9M$. The Killing horizon has spherical topology consisting of an inner and outer spherical surface. Center panel: $|a|=M$. The inner and outer Killing horizons merge at the poles. Right panel: $|a|=1.1M$. The topology of the Killing horizon is toroidal. In all cases, the central object is a naked singularity located at the Killing horizon. For all values of the spin, the origin is likewise singular.}
\label{topochange11}
\end{figure*}

In this paper, I only discuss the various metrics in the spin range $|a| \leq M$, because the Kerr background itself is pathological outside of this range (see the discussion in the next section). For completeness, however, I investigate the nature of the central object in the MK metric for values of the spin $|a|>M$ observing another change in the topology of the Killing horizon. If $|a|>M$, Eq.~\eqref{Killing_plane} has two solutions if $\epsilon_3\leq0$ and no solutions otherwise, while Eq.~\eqref{Killing_poles} has two solutions if $\epsilon_3<\epsilon_3^{\rm Kil-pol}$, one solution if $\epsilon_3=\epsilon_3^{\rm Kil-pol}$, and no solution otherwise. Solving Eq.~\eqref{horizon1} at the remaining polar angles, I find that the inner and outer Killing horizons have spherical topology if $\epsilon_3\leq\epsilon_3^{\rm Kil-pol}$ as already found in Ref.~\cite{MKmetric}. If $\epsilon_3=\epsilon_3^{\rm Kil-pol}$, the inner and outer Killing horizons coincide at the poles. If $\epsilon_3>\epsilon_3^{\rm Kil-pol}$, the Killing horizon has toroidal topology and is centered in the equatorial plane. In Fig.~\ref{topochange11}, I plot several examples of the different topologies. Such changes in topology are similar to the ones reported in Ref.~\cite{topology}. 

Finally, I investigate the nature of the central object in the MK metric when it is treated as a small perturbation of the Kerr metric in the sense of Eq.~(\ref{eq:gen-form}). In this case, I can calculate the location of the null surface from Eq.~\eqref{horeqpert1}, which, in this case, turns out to be an event horizon. Note that the metric element $h_{rr}^{\rm MK}$ in expression Eq.~\eqref{MKlinearized} diverges as $r \to r_+$; however, $h^{rr}_{\rm MK} = g^{rr}_{\rm K}g^{rr}_{\rm K}h_{rr}^{\rm MK}$ is well behaved there, since $g^{rr}_{\rm K} = \Delta/\Sigma$, canceling the factor $\Delta^{-2}$. Since this expression is still valid for all spin values $|a|\leq M$, I expect the event horizon to be located at the radius
\be
r_H = r_+(1 + \lambda\epsilon_3),
\ee
where $r_+$ is the Kerr horizon given by Eq.~\eqref{kerrhorizon} and where $\lambda$ is the amount, by which the horizon is modified relative to the Kerr horizon. Inserting this expression into Eq.~\eqref{horeqpert1} and linearizing in the parameter $\epsilon_3$, I obtain the equation
\be
\lambda = -\frac{ \epsilon_3 a^2 M^3 r_+ \sin^2\theta }{ 2\sqrt{M^2-a^2} \left( 2M r_+ - a^2\sin^2\theta \right)^2 }.
\ee
The event horizon is then located at the radius
\be
r_H = r_+ \left[ 1 - \frac{ \epsilon_3 a^2 M^3 \sin^2\theta }{ 2\sqrt{M^2-a^2} \left( 2M r_+ - a^2\sin^2\theta \right)^2 } \right].
\ee
The Kretschmann scalar of the linearized MK metric remains finite at all radii $r>0$, and the central object is a black hole for both positive and negative values of the parameter $\epsilon_3$.

\section{Lorentz Violations, Closed Timelike Curves, and Regions of Validity}
\label{sec:vnht}

Each of the metrics discussed in Sec.~\ref{sec:pdKM} may harbor regions of space where the Lorentzian symmetry is broken or which contain closed timelike curves. In Sec.~\ref{asy-flat}, I studied the asymptotic structure of the metrics at spatial infinity, but this does not guarantee that they retain their Lorentzian signatures close to the central objects. Since I already showed that the metrics have Lorentzian signature at spatial infinity since they are asymptotically flat, their signature must change sign close but outside their central objects. In that case, their determinants must vanish somewhere outside of their central objects, making them singular. If such regions exist, the metric can only describe physical processes outside of them. In the following, I analyze the properties of the five metrics in this regard.

\subsection{Lorentz violations}

When I say that a metric is of Lorentzian signature, I mean that 
\be
{\rm{det}} \left(g_{\mu \nu}\right) < 0\,. 
\ee
In the case of the Minkowski metric, the determinant is simply $-1$. For the Kerr metric, the determinant is
\begin{align}
{\rm{det}} \left(g^{\Kerr}_{\mu \nu}\right) = -  \Sigma^{2} \sin^{2}{\theta}\,,
\end{align}
which is clearly negative definite everywhere outside the singularity (as well as the poles, $\theta=0,\pi$). Similarly, I will study whether the metrics proposed in Sec.~\ref{sec:pdKM} remain of Lorentzian signature everywhere outside their central objects.

\subsection{Closed timelike curves}

In accordance with the no-hair theorem, the exterior domain of the Kerr metric, i.e., the domain outside the event horizon, is causally well-behaved and free of closed timelike curves if $|a|\leq M$~\cite{Carter68}. This follows from the fact that, for constant times $t$, the hypersurfaces $(r,\theta,\phi)$ are always spacelike. This comes about because for constant $(t,\phi)$, the two-dimensional metric induced on surfaces $(r,\theta)$ is positive definite and 
\begin{equation}
\zeta_\mu^{(\phi)} \zeta^\mu_{(\phi)} = g_{\phi\phi} > 0\,,
\end{equation}
where $\zeta^\mu_{(\phi)}=(0,0,0,1)$ is the axial Killing vector~\cite{Carter68,Carter73}. If $|a|>M$, the event horizon disappears and a naked singularity emerges. In this case, causality is violated everywhere, because any event in that spacetime can be connected to any other event by both a future and a past directed timelike curve~\cite{Carter68,Carter73} (see, also, Ref.~\cite{CdF82}).

Closed timelike curves may exist if the metric is no longer Lorentzian or if $\zeta_\mu \zeta^\mu < 0$. In the latter case, e.g. circles with $t={\rm const}$, $r={\rm const}$, $\theta={\rm const}$ are closed timelike curves, because the vector $\zeta^\mu$ is timelike, which can be thought of as the ``tipping over'' of light cones. In the Kerr spacetime with $|a|\leq M$, closed timelike curves exist inside the (inner) event horizon $r_{-}$~\cite{Carter68}. Other examples of spacetimes with closed timelike curves include the van Stockum spacetime~\cite{vanStockum}, the G\"odel universe~\cite{Godel49}, and the Gott cosmic string~\cite{Gott91}.

If a given metric contains a pathological region outside of the central object that I identified in the previous section, I also calculate the largest radius at which the metric becomes pathological as a function of both the spin and the parameter $\zeta$. This outermost radius serves as an indicator that a cutoff radius has to be introduced which properly excises the pathological region.

In this section, I discuss the determinants of the QK, BK, and MGBK metrics to first order in the deviation parameters, while I discuss the full determinants of the MN and MK metrics. If I likewise analyze the determinants of the QK, BK, and MGBK metrics to all orders in the deviation parameters, I find that, if regions where Lorentz symmetry is violated are present, their locations are slightly modified, but qualitatively the same. My findings for the regions containing closed timelike curves are unchanged because of the linear form of the $(\phi,\phi)$ elements in these metrics.

\subsubsection{QK metric}

\begin{figure}[ht]
\begin{center}
\psfig{figure=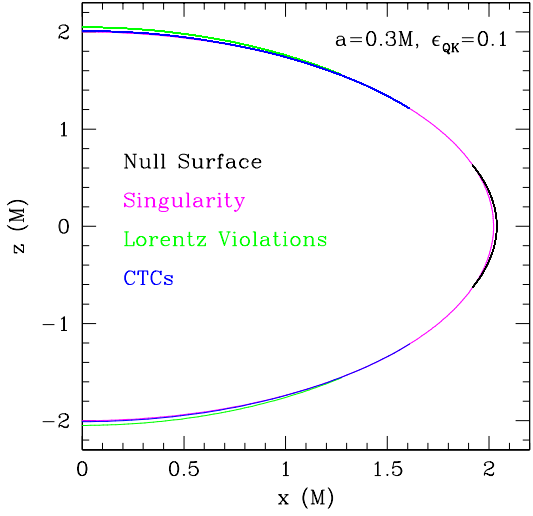,height=3.2in}
\psfig{figure=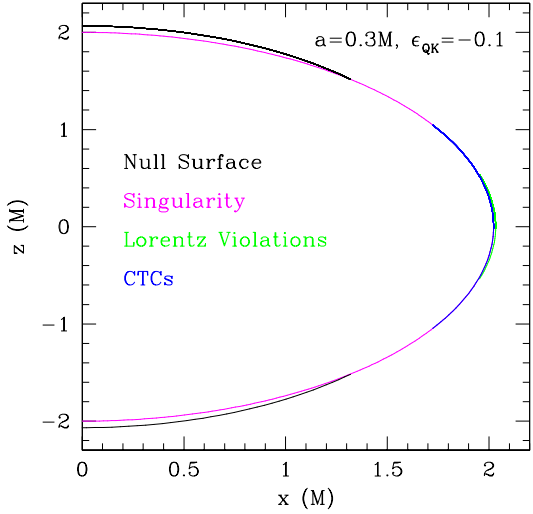,height=3.2in}
\end{center}
\caption{Null surface and regions with Lorentz violations and closed timelike curves (denoted ``CTCs") in the QK metric for $a=0.3M$. Top panel: $\epsilon_\QK=0.1$; Lorentz violations and closed timelike curves occur around the poles. Bottom panel: $\epsilon_\QK=-0.1$; Lorentz violations and closed timelike curves occur near the equatorial plane.}
\label{GBvalidity}
\end{figure}

To linear order in the parameter $\zeta_{\rm QK}$, the determinant of the quasi-Kerr metric is given by the expression
\ba
\det \left( g_{\mu\nu}^\QK \right) = && - \sin^2 \theta \left\{ \Sigma^2 - \frac{5\epsilon_\QK r^3}{16M^2} (1 + 3\cos 2\theta) \nn \right. \\
&& \bigg[ 2M (2M^2 -3Mr -3r^2)  \nn \\
&& \left. + 3r(r^2 - 2M^2) \ln \left( \frac{r}{r-2M} \right) \bigg] \right\}.
\ea
If $\epsilon_\QK \neq 0$, this determinant changes sign outside of the singularity at $r=2M$, and hence the QK metric becomes non-Lorentzian. In addition, regions with closed timelike curves exist outside of the singularity. Regions of Lorentz violations and closed timelike curves are present around the poles if $\epsilon_\QK>0$ and near the equatorial plane if $\epsilon_\QK<0$. In Fig.~\ref{GBvalidity}, I plot these regions for the case $a=0.3M$ and $\epsilon_\QK = \pm0.1$.

\begin{figure}[ht]
\begin{center}
\psfig{figure=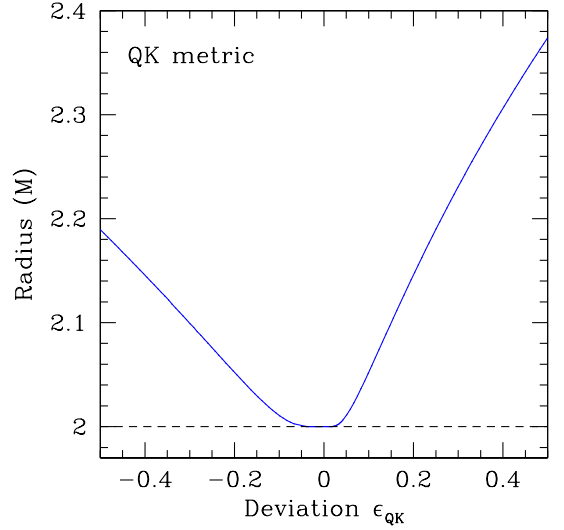,height=3.2in}
\end{center}
\caption{Outermost radius, at which a pathology occurs in the QK metric, as a function of the deviation parameter $\epsilon_{\rm QK}$. This radius depends only very weakly on the value of the spin. Note that I extrapolated this radius to include larger values of the deviation parameter; in this case, the location of the outermost radius is only approximate. In the range of the parameter $-0.5\leq\epsilon_{\rm QK}\leq0.5$, the outermost radius lies well inside of the ISCO radius (see Ref.~\cite{PaperI}). The dashed line corresponds to the naked singularity located at the radius $r=2M$.}
\label{QKcutoff}
\end{figure}

In Fig.~\ref{QKcutoff}, I plot the outermost radius, at which a Lorentz violation occurs, as a function of the deviation parameter. At each radius, pathological regions can only lie on or inside of this radius. For positive values of the parameter $\epsilon_{\rm QK}$, this outermost radius of the Lorentz-violating region is located at the poles, while for negative values of the parameter $\epsilon_{\rm QK}$, the outermost radius is located in the equatorial plane. This radius depends only very weakly on the value of the spin. In Fig.~\ref{QKcutoff}, I extrapolated the location of the outermost radius to include larger values of the parameter $\epsilon_{\rm QK}$; this location is only approximate. In the range of the parameter $-0.5\leq\epsilon_{\rm QK}\leq0.5$, the outermost radius lies well inside of the ISCO radius (see Ref.~\cite{PaperI}).

\subsubsection{BK metric}

\begin{figure}[ht]
\begin{center}
\psfig{figure=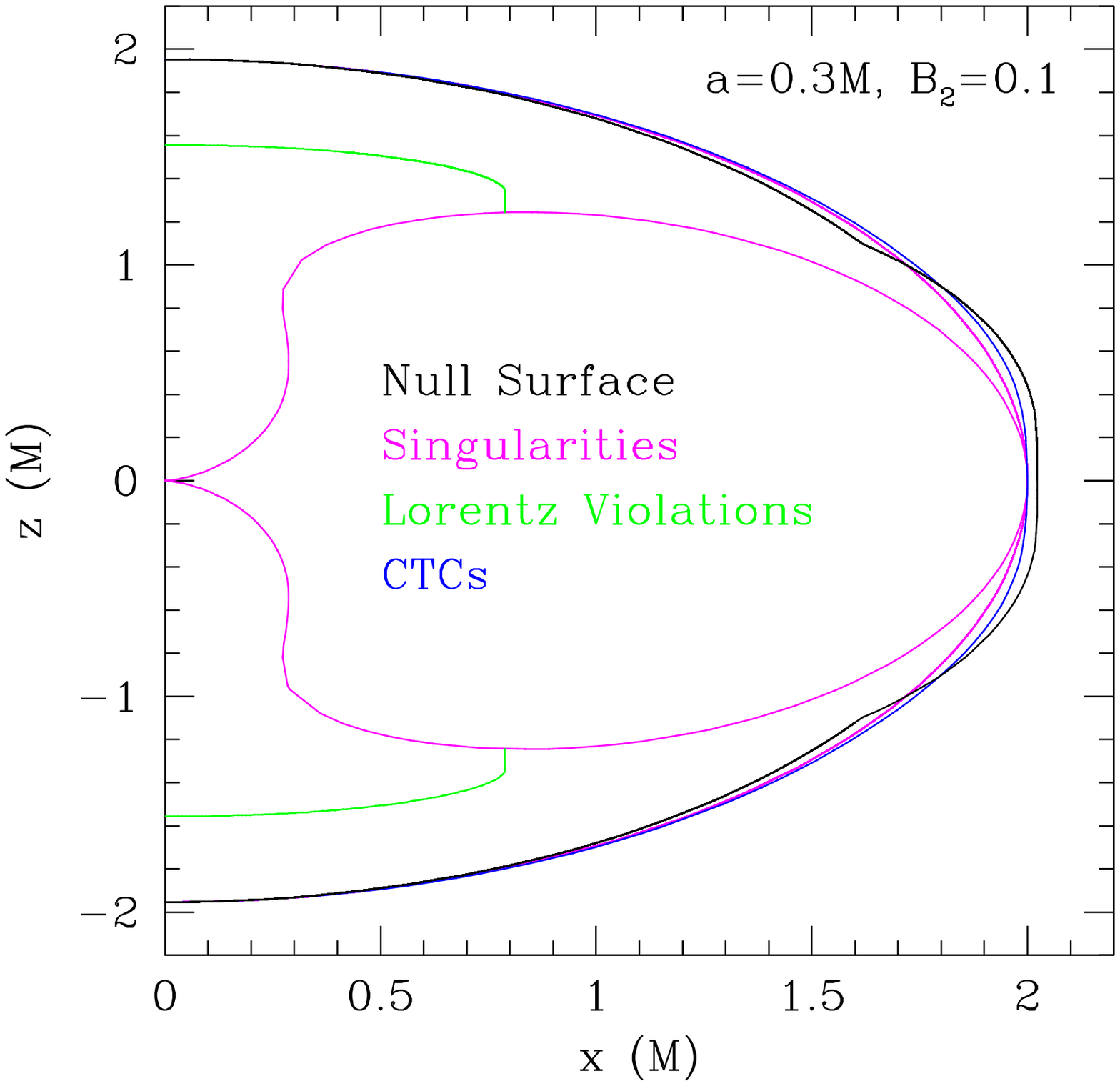,height=3.2in}
\psfig{figure=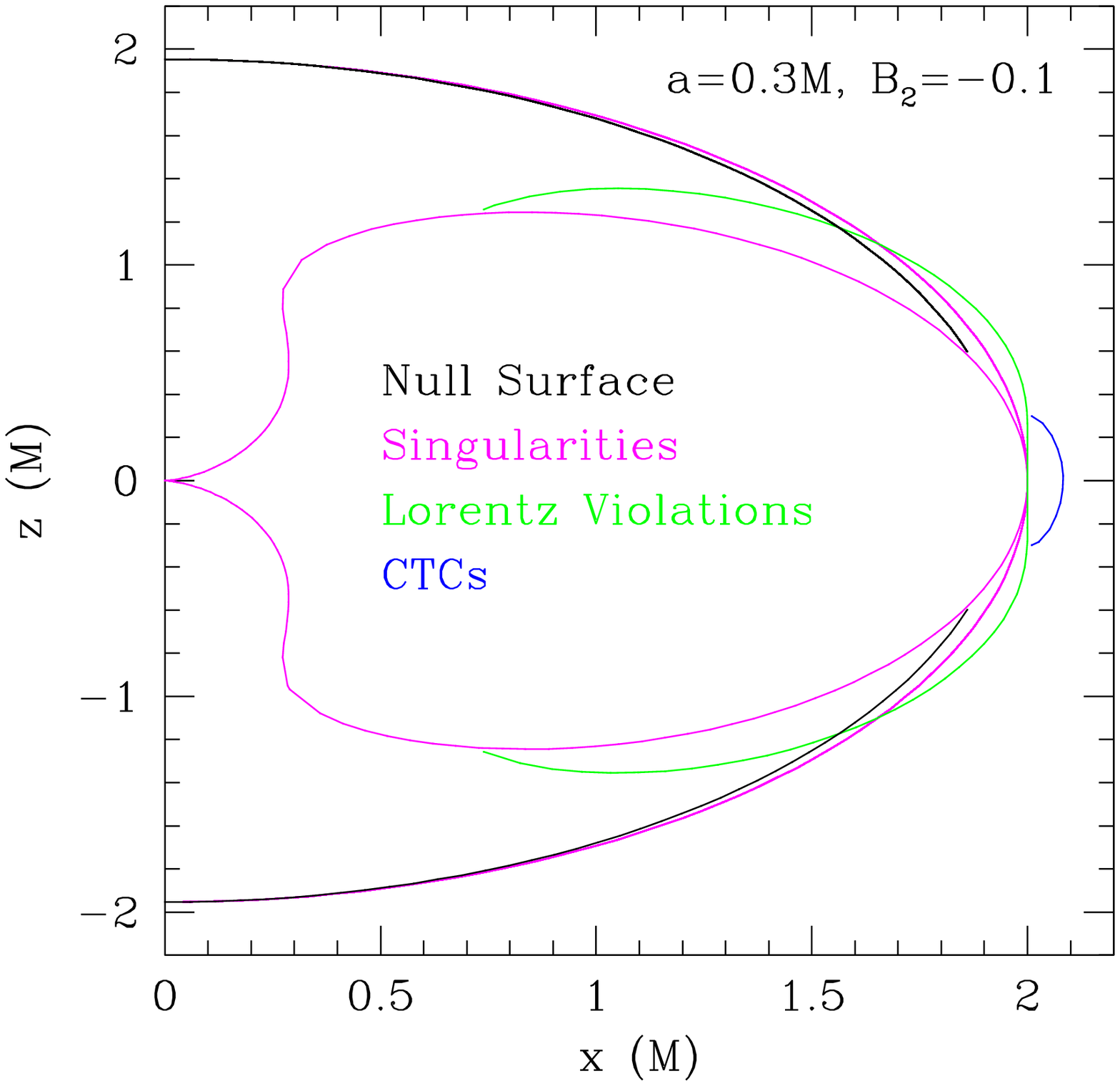,height=3.2in}
\end{center}
\caption{Null surface and regions with Lorentz violations and closed timelike curves in the BK metric for $a=0.3M$ and $B_2=\pm0.1$. The null surface is closed if $B_2=0.1$ and open if $B_2=-0.1$. In both cases, Lorentz violations or closed timelike curves occur outside of the null surface and the outermost singularity.}
\label{BKvalidity}
\end{figure}

In the case of the BK metric, to linear order in the parameter $\zeta_{\rm BK}$, the determinant is given by the expression
\ba
{\rm{det}} \left(g_{\mu \nu}^\BK\right) = -\Sigma^2 \sin^2\theta \left[ 1 - 4(\psi_1-\gamma_1) \right] \,, 
\ea
which becomes non-Lorentzian when
\begin{equation}
\psi_1-\gamma_1 > \frac{1}{4} \;.
\end{equation}

For this spacetime, Lorentz violations and closed timelike curves occur outside of the null surface and the outermost singularity $r_{1,+}$ around the equatorial plane if $B_2\neq0$. Additional regions of Lorentz violation lie inside the null surface. In Fig.~\ref{BKvalidity}, I plot these regions for $a=0.3M$ and $B_2=\pm0.1$.

\begin{figure}[ht]
\begin{center}
\psfig{figure=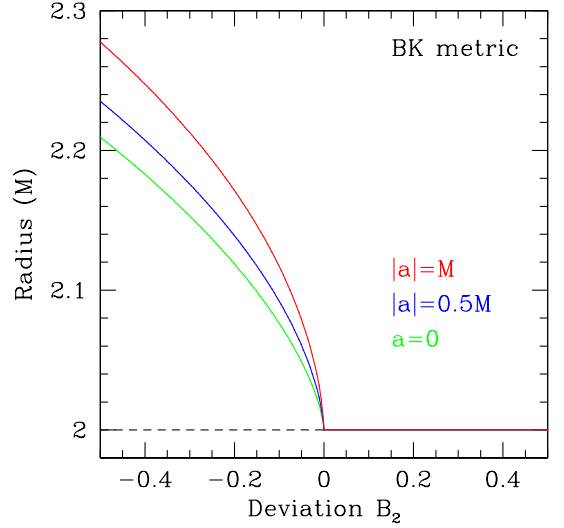,height=3.2in}
\end{center}
\caption{Outermost radius, at which a pathology occurs in the BK metric, as a function of the deviation parameter $B_2$ for several values of the spin $a$. Note that I extrapolated this radius to include larger negative values of the deviation parameter; in this case, the location of the outermost radius is only approximate. The dashed line corresponds to the equatorial singularity located at $r_{1,+}=2M$, which is independent of the spin.}
\label{BKcutoff}
\end{figure}

In Fig.~\ref{BKcutoff}, I plot the outermost equatorial radius, at which the BK metric becomes pathological, as a function of the deviation parameter $B_2$ for several values of the spin. For positive values of the parameter $B_2$, this radius is equal to the radius of the singularity $r_{1,+}$ given by Eq.~\eqref{BK_sing1}. For negative values of the parameter $B_2$, this radius is equal to the radius, at which a closed timelike curve is located. For larger negative values of the parameter $B_2$, I extrapolated the location of the outermost radius; this location is only approximate. At each radius, pathological regions can only lie on or inside of this radius.

\subsubsection{MN metric}

\begin{figure}[ht]
\begin{center}
\psfig{figure=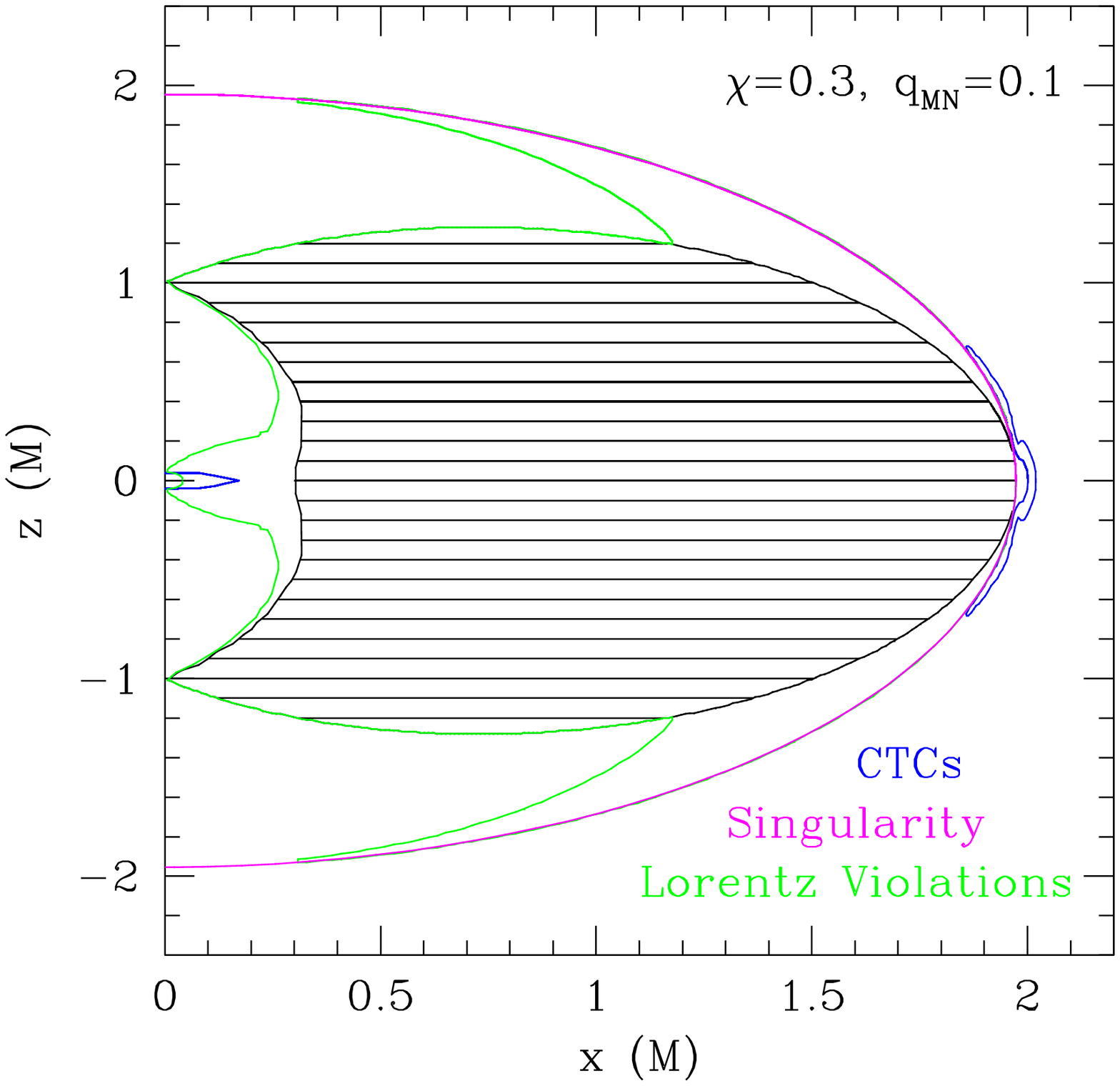,height=3.2in}
\psfig{figure=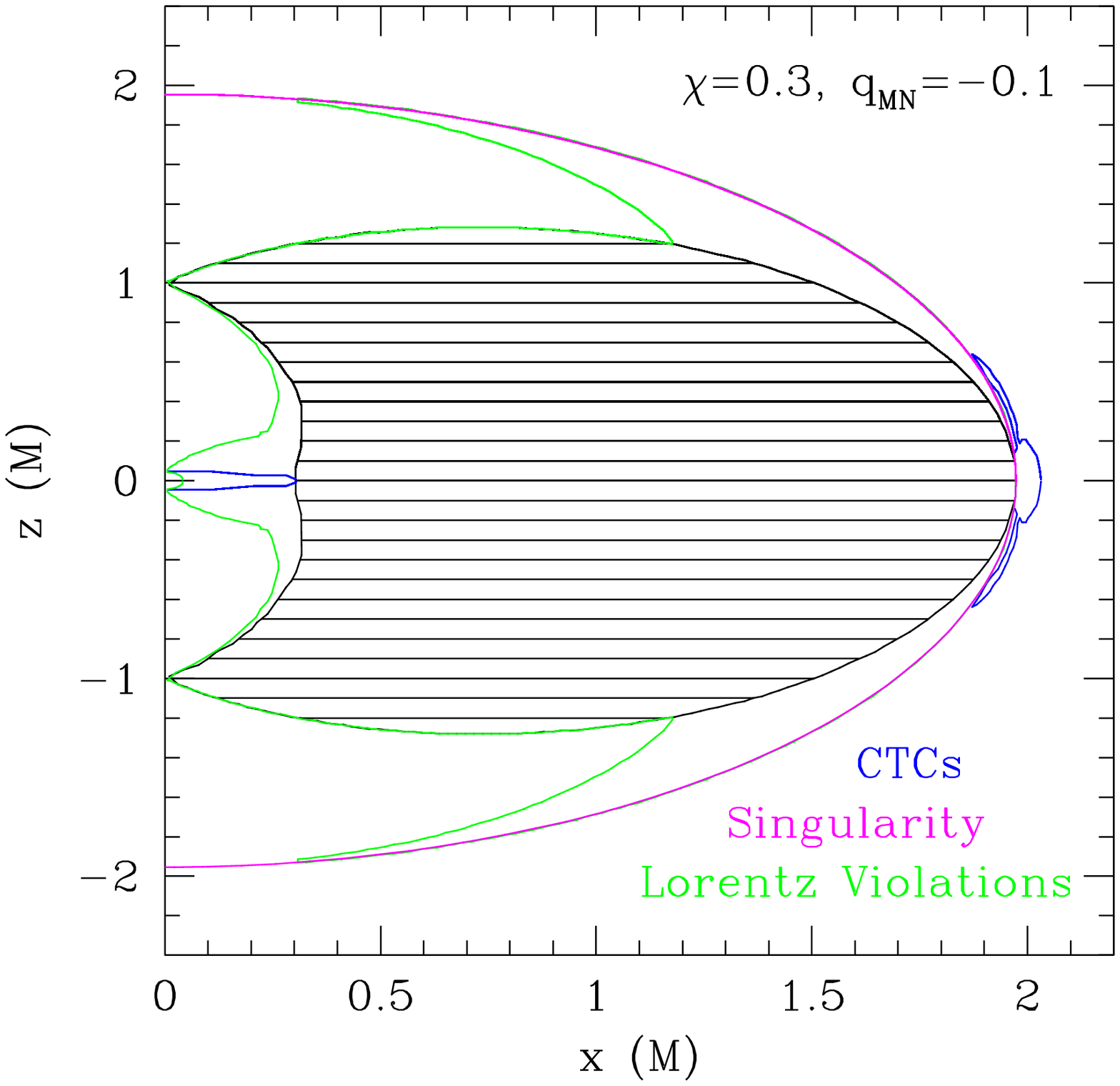,height=3.2in}
\end{center}
\caption{Singularity and regions with Lorentz violations and closed timelike curves in the MN metric for $\chi=0.3$ and $q_\MN=\pm0.1$. Lorentz violations lie inside and closed timelike curves occur both inside and outside of the (singular) boundary that coincides to the event horizon of a Kerr black hole with the same value of the spin. In the black shaded region the MN metric is imaginary.}
\label{MNvalidity}
\end{figure}

The MN metric is of a much more complex form, and I will not write its determinant here explicitly. Gair et al.~\cite{Gair08}, Brink~\cite{Brink08}, and Bambi~and~Lukes-Gerakopoulos~\cite{BLG13} analyzed this metric in cylindrical coordinates without the rescaling of Sec.~\ref{sec:pdKM}. They showed the existence of regions with closed timelike curves around the singularity. In the form used in this paper, if $q_\MN \neq 0$, closed timelike curves occur around the equatorial plane outside and inside of the surface that numerically coincides with the Kerr event horizon. In the majority of this region, the MN metric is imaginary. In Fig.~\ref{MNvalidity}, I plot these regions for $|\chi|=0.3$ and $q_\MN = \pm 0.1$.

\begin{figure}[ht]
\begin{center}
\psfig{figure=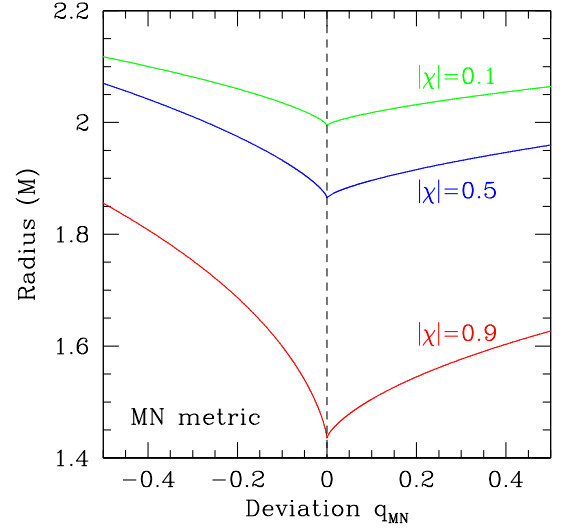,height=3.2in}
\end{center}
\caption{Outermost radius, at which a pathology occurs in the MN metric, as a function of the deviation parameter $q_{\rm MN}$ for several values of the spin $a=\chi M$. The dashed line corresponds to a Kerr black hole.}
\label{MNcutoff}
\end{figure}

In Fig.~\ref{MNcutoff}, I plot the outermost equatorial radius, at which a closed timelike curve occurs in the equatorial plane of the MN metric, as a function of the deviation parameter $q_{\rm MN}$ for several values of the spin. At each radius, pathological regions can only lie on or inside of this radius.

If I study the determinant and the $(\phi,\phi)$ element of the MN metric to linear order in the deviation parameter, I find that the locations of the regions containing Lorentz violations or closed timelike curves are shifted slightly.

\subsubsection{MK metric}

Evaluating the determinant of the MK metric,
\ba
\det\left(g_{\mu\nu}^{\rm MK}\right) = && -\frac{\sin^2\theta}{64\Sigma^2} \big[ 3a^4+8a^2r^2+8r^4+8\epsilon_3 M^3 r \nn \\
&& + 4a^2(2r^2+a^2)\cos{2\theta} + a^4 \cos{4\theta} \big]^2,
\ea
I notice that it is negative semidefinite and vanishes at two radii $r_{S,\pm}(\theta)$ if $\epsilon_3<-4r_+$. These radii coincide with the locations of the Killing horizons as found in the previous section, because of the relation
\be
\det\left(g_{\mu\nu}^{\rm MK}\right) \propto \left[ g_{tt}^{\rm MK}g_{\phi\phi}^{\rm MK} - \left(g_{t\phi}^{\rm MK}\right)^2\right],
\ee
see Eq.~\eqref{horizon1}. Therefore, the MK metric does not contain any Lorentz-violating regions.

\begin{figure}[h]
\begin{center}
\psfig{figure=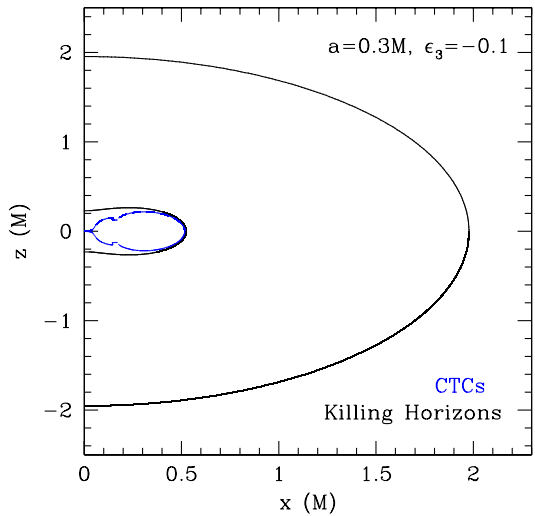,height=3.2in}
\end{center}
\caption{Inner and outer Killing horizons and regions with closed timelike curves in the MK metric for $a=0.3M$ and $\epsilon_3=-0.1$. The metric is regular outside of the outer Killing horizon, which is singular and located just outside of the null surface. Regions with closed timelike curves occur only inside the inner Killing horizon.}
\label{MKvalidity}
\end{figure}

From the $g_{\phi\phi}^{\rm MK}$ element in Eq.~(\ref{metricMK}), I can see that for given values of the radius and the spin closed timelike curves only occur for values of the parameter
\be
\epsilon_3 < - \frac{r M^3 \Delta}{a^2 \Sigma^2}\leq0.
\label{MKctc}
\ee
No closed timelike curves exist if $\epsilon_3\geq0$. In Fig.~\ref{MKvalidity}, I plot this region for $a=0.3M$ and $\epsilon_3=-0.1$. I can see that the region containing closed timelike curves is located inside of the inner Killing horizon in analogy to the Kerr metric, where closed timelike curves lie inside the inner event horizon. The values of the deviation parameter in Eq.~(\ref{MKctc}) have an upper bound at
\be
\epsilon_3^{\rm CTC} = - \frac{r^3 [ r^3 + a^2 (r+2M) ]}{a^2(r+2M)}
\ee
corresponding to the polar angle $\theta=\pi/2$. Solving Eq.~(\ref{grrzero}) in the equatorial plane for the parameter $\epsilon_3$, I find
\be
\epsilon_3^{\rm hor} = - \frac{r^3 \Delta}{a^2 M^3}.
\ee
It is easy to see that $\epsilon_3^{\rm CTC}<\epsilon_3^{\rm hor}$ for all values of the radius and the spin. Consequently, at least in the equatorial plane, closed timelike curves always lie inside the inner Killing horizon. Generally, I find numerically that the region containing closed timelike curves is always located inside of the outer Killing horizon. Therefore, the MK metric does not contain any closed timelike curves outside of the central object (see also Ref.~\cite{MKmetric}).

For the determinants and $(\phi,\phi)$ elements of MK metric expanded to linear order in the deviation parameter, I find that Lorentz violations occur inside of the central object, while the locations of the regions containing closed timelike curves are shifted slightly (still remaining inside of the central object).

\subsubsection{MGBK metric}

\begin{figure}[h]
\begin{center}
\psfig{figure=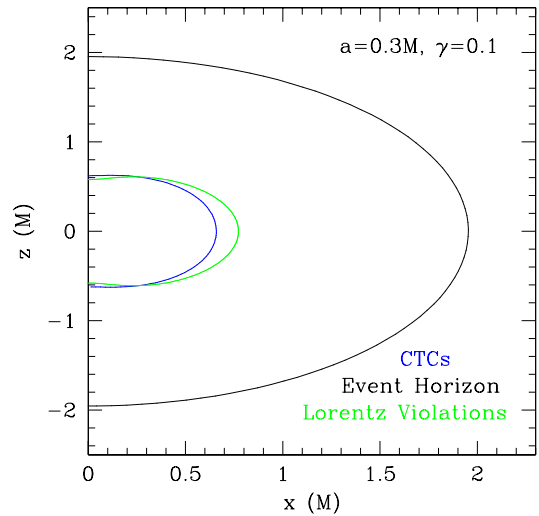,height=3.2in}
\end{center}
\caption{Event horizon and regions with Lorentz violations and closed timelike curves in the MGBK metric for $a=0.3M$ and $\gamma_{1,2}=\gamma_{3,1}=\gamma_{3,3}=\gamma_{4,2}=0.1$. The metric is regular outside of the event horizon, and regions with Lorentz violations and closed timelike curves occur only inside the event horizon.}
\label{MGBKvalidity}
\end{figure}

In the case of the MGBK metric, the determinant to linear order in the deviation parameters is, again, a very long expression, and I will not write it explicitly. The metric is regular everywhere outside of the event horizon, and regions with Lorentz violations and closed timelike curves occur only inside the event horizon as long as $|\bar{\gamma}_{i,n}|\leq0.1$. In Fig.~\ref{MGBKvalidity}, I plot these regions for $a=0.3M$ and $\gamma_{1,2}=\gamma_{3,1}=\gamma_{3,3}=\gamma_{4,2}=0.1$. Violations can occur outside of the horizon when $|\bar{\gamma}_{i,n}|>0.1$.

In Table~1, I summarize the properties and pathologies of the various metrics.

\begin{table*}[ht]
\begin{center}
\footnotesize
\begin{tabular}{lccccccl}
\multicolumn{8}{c}{{\bf Table 1.} Properties of Kerr-like Metrics}\\
\hline \hline

Metric & Deviation & Ricci & Petrov & Central & Lorentz              & Closed timelike  & ~~~~~~~~~~~~~~~~~~~Reference \\
       & type      & flat  & type   & object  & violations${\rm ^a}$ & curves${\rm ^a}$ & \\
\hline
QK & Linear${\rm ^b}$ & Yes & I & Naked singularity & Yes & Yes & Glampedakis and Babak (2006), Ref.~\cite{GB06}\\
BK & Linear${\rm ^b}$ & No${\rm ^c}$ & I & Naked singularity & Yes & Yes & Vigeland and Hughes (2010), Ref.~\cite{VH10}\\
MN & Nonlinear & Yes & I & Naked singularity & No & Yes & Manko and Novikov (1992), Ref.~\cite{MN92}\\
MK & Nonlinear & No & I & Naked singularity${\rm ^d}$ & No & No & Johannsen and Psaltis (2011), Ref.~\cite{MKmetric}\\
MGBK & Linear${\rm ^b}$ & No & D & Black hole & No & No & Vigeland, Yunes, and Stein (2011), Ref.~\cite{Vigeland:2011ji}\\
\hline
\end{tabular}
\end{center}
\footnotetext{Lorentz violations and closed timelike curves refer to pathologies outside of the central objects.}
\footnotetext{I analyzed the properties of the spacetimes with a linear deviation from the Kerr metric only for small values of the respective deviation parameters, c.f. Eq.~\eqref{eq:gen-form}. These properties may be different for larger deviations.}
\footnotetext{Ricci flat if $a=0$.}
\footnotetext{Black hole for small deviations at linear order.}
\caption{Properties of the Kerr-like metrics: The quasi-Kerr metric (QK) modifies the quadrupole moment of the Kerr metric after the quadrupole moment of the Hartle-Thorne metric that is a solution of the vacuum Einstein equations for small values of the spin and the quadrupolar deviation. The bumpy Kerr metric (BK) augments the Kerr metric with arbitrary small multipolar distortions, with a functional form chosen to insure that it remains a vacuum solution of the linearized Einstein equations in the zero spin limit. The metric proposed by Manko and Novikov (MN) is a Ricci flat generalization of the Kerr metric with arbitrary multipole moments; it is an exact solution of the Einstein equations. The metric proposed by Johannsen and Psaltis (MK) deviates from the Kerr metric in nonlinear form and introduces a set of polynomial modifications to the spacetime. The modified-gravity bumpy Kerr metric (MGBK) is an extension of the BK metric designed to admit an approximate third constant of the motion.}
\end{table*}

\section{Discussion}
\label{sec:discussions}

In this paper, I compiled several parametric frameworks of metrics that deviate from the Kerr solution. Due to the special nature of the Kerr metric in general relativity as a consequence of the no-hair theorem, all such parametric spacetimes have to differ in at least one of the properties of the Kerr metric, which I analyzed in detail.

The MN metric as well as the QK and BK metrics in the appropriate limits are vacuum solutions in general relativity, while the MK and MGBK metrics are not. The MGBK metric admits three constants of the motion and is of approximate Petrov type D, while the other four metrics generally admit only two constants of the motion and are of Petrov type I. All five metrics are asymptotically flat, which I showed explicitly for the QK and MN metrics. The MN metric, however, requires an appropriate coordinate transformation and a rescaling of the mass. The QK, BK, and MGBK metrics are designed as linear deviations from the Kerr metric, while the MN and MK metrics are nonlinear parametric deviations.

I described the calculation of event horizons in stationary, axisymmetric, asymptotically flat metrics using numerical relativity techniques. I applied this approach to the spacetimes I studied in this paper and showed that the QK and BK metrics harbor naked singularities. The MN metric likewise describes a naked singularity \cite{MN92}, while the MGBK metric harbors a black hole. I also showed that the MK metric contains a naked singularity of spherical topology for values of the parameter $\epsilon_3<\epsilon_3^{\rm max-eq}$ and in the form of two disjoint spherical surfaces if $\epsilon_3\geq\epsilon_3^{\rm max-eq}$, see Eq.~\eqref{ep3maxeq}. I determined this bound analytically. If treated as a small perturbation of the Kerr metric, the MK metric likewise describes a black hole for nonzero values of the deviation parameter and all values of the spin. I also identified regions with Lorentz violations or closed timelike curves outside of the central objects of the QK, BK, and MN metrics, while the MK and MGBK metrics are free of such pathologies exterior to the naked singularity and event horizon, respectively.

All of these metrics can be used for astrophysical tests of the no-hair theorem in either the electromagnetic or gravitational-wave spectrum. However, the existence of pathologies impacts the utility of some of these metrics for such tests. In order to shield the adverse effects of the pathological regions outside of the central object from distant observers, a cutoff radius needs to be defined as an inner boundary of the exterior spacetime, where a given metric is free of any unphysical behavior. Such a cutoff acts as an artificial horizon, which ``captures" any matter or radiation that passes through it, i.e., any particle entering the domain inside the cutoff leaves the exterior spacetime permanently. 

Tests of the no-hair theorem with both electromagnetic and EMRI observations probe radii that are comparable to the location of the ISCO. In many accretion disk models, the ISCO often marks the inner disk edge, while future gravitational-wave detectors will be most sensitive to EMRIs occurring at radii roughly in the range between the innermost stable orbit and $10-20M$. Therefore, it is critical for such tests that this region can be properly modeled. In the Kerr metric the location of the ISCO decreases for increasing values of the spin and merges with the coordinate location of the event horizon in the limit $a\rightarrow M$. In parametrically deformed Kerr spacetimes, the location of the ISCO generally depends on the spin and the deviation parameters (see, e.g., Ref.~\cite{PaperI}). Thus, if a cutoff radius has to be introduced in a given metric, the metric can only be used for tests of the no-hair theorem for values of the spin and the deviation parameters for which the ISCO lies outside of the cutoff radius (see the discussion in Refs.~\cite{PaperI,MKmetric}). This implies that the BK and MN metrics (as well as the QK metric by construction) can only be used for moderately spinning black holes as long as the ISCO lies outside of the cutoff radius. For these three spacetimes, I calculated the outermost radius, at which the spacetime is no longer well behaved, as a function of the spin and the deviation parameter. In each case, a cutoff radius has to be introduced so that the pathological regions are excised. In practical applications, it is often convenient to choose a spherical cutoff, which is located just outside of this outermost radius. However, more sophisticated choices of a cutoff radius which depend on the polar angle $\theta$ are also possible.

The MK metric is regular everywhere outside of the naked singularity located at the Killing horizon for all values of the spin $|a| \leq M$. In the region of the parameter space where an ISCO exists, the ISCO is always located outside of this surface and, as shown in Ref.~\cite{MKmetric}, the ISCO only coincides with the Killing horizon in the Kerr case if $a=M$ and $\epsilon_3=0$. Therefore, if needed, a cutoff can always be chosen at a radius between the naked singularity and the ISCO (a similar property also holds for the circular photon orbit; see Ref.~\cite{MKmetric}). 

In several settings of accretion flows around black holes, the presence of an event horizon affects only marginally the characteristic of the flow itself or its observational appearance. Inside the ISCO (i.e., in the plunging region) the plasma attains highly supersonic inward velocities and gets causally disconnected from the hydrodynamics of the material outside the ISCO. Imposing a cutoff, therefore, only has a very minor effect, and the MK metric provides a useful framework for electromagnetic tests of the no-hair theorem for all values of the spin $|a| \leq M$ \cite{MKmetric}. In other situations, however, the presence and properties of the event horizon modifies significantly the accretion flow outside the ISCO. This is especially true for spinning black holes with significant magnetic flux threading the horizon, potentially arresting the flow, and launching powerful jets. In such cases, (e.g., in fully relativistic magnetohydrodynamic simulations), more care must be taken to ensure that fields and material in the simulation do not come into contact with a singularity. I will analyze the utility of the MK metric in this scenario in a future paper.

The MGBK metric is regular everywhere outside of the event horizon as long as it remains a small perturbation of the Kerr metric. In this regime, this metric can likewise be used for electromagnetic tests of the no-hair theorem. However, since this metric is of Petrov type D, it admits an approximate third constant of the motion, a Carter constant. This property is useful to construct model waveforms, and the MGBK metric is, therefore, a useful framework for gravitational-wave tests of the no-hair theorem \cite{Vigeland:2011ji,Gair:2011ym}.

\acknowledgments
I thank S. Hughes, D. Psaltis, S. Vigeland, and N. Yunes for many useful comments as well as V. Manko and J. McKinney for useful conversations. I was supported by a CITA National Fellowship at the University of Waterloo and by the National Science Foundation under Contract No. NSF 0746549 while at the University of Arizona. Research at Perimeter Institute is supported by the Government of Canada through Industry Canada and by the Province of Ontario through the Ministry of Research and Innovation.

\appendix
\section{Coordinate transformations in the MN spacetime}

I begin with the line element as given by Ref.~\cite{MN92} (see, also,~\cite{Gair08})
\begin{align}
ds_{\MN}^2 &= -f_{\MN}(dt-\omega d\phi)^2 + k^2 f_{\MN}^{-1} e^{2\Gamma} (x^2 - y^2) 
\nonumber \\
&\times
\left( \frac{ dx^2 }{ x^2 -1 } + \frac{ dy^2 }{ 1-y^2 } \right) + k^2 f_{\MN}^{-1} (x^2-1)(1-y^2)d\phi^2,
\end{align}
where
\allowdisplaybreaks[1]
\begin{align}
f_{\MN} &= e^{2\Psi} \frac{A}{B}, \nonumber \\
\omega &= 2k e^{-2\Psi} \frac{C}{A} - \frac{ 4k\alpha }{ 1-\alpha^2 }, \nonumber \\
e^{2\Gamma} &= e^{ 2\Gamma'} \frac{ A }{ (x^2-1)(1-\alpha^2)^2 }, \nonumber \\
A &= (x^2-1)(1+uv)^2 - (1-y^2)(v-u)^2, \nonumber \\
B &= [x+1+(x-1)uv]^2 + [(1+y)u + (1-y)v]^2, \nonumber \\
C &= (x^2-1)(1+uv)[v-u-y(u+v)] 
\nonumber \\
&+ (1-y^2)(v-u)[1+uv+x(1-uv)], \nonumber \\
\Psi &= \beta \frac{P_2}{R^3}, \nonumber \\
\Gamma' &= \frac{1}{2} \ln \frac{x^2-1}{x^2-y^2} + \frac{9\beta^2}{6R^6} (P_3^2 - P_2^2) 
\nonumber \\
&+ \beta \sum_{l=0}^2 \left[ \frac{ x-y+(-1)^{2-l}(x+y) }{ R^{l+1} }P_l - 2 \right], \nonumber \\
u &= -\alpha~{\rm exp}\left[- 2\beta \left( -1 + \sum_{l=0}^2 \frac{ (x-y)P_l }{ R^{l+1} } \right) \right], \nonumber \\
v &= \alpha~{\rm exp}\left[ 2\beta \left( 1 + \sum_{l=0}^2 \frac{ (-1)^{3-l} (x+y)P_l }{ R^{l+1} } \right) \right], \nonumber \\
R &\equiv \sqrt{ x^2 + y^2 -1 }, \nonumber \\
P_n &\equiv P_n\left(\frac{xy}{R}\right),~~~P_n(x) = \frac{1}{2^n n!} \left(\frac{d}{dx}\right)^n (x^2-1)^n.
\end{align}
In these equations, $k$, $\alpha$, and $\beta$ are free parameters, which determine the set of multipole moments of the spacetime. Note that there is a minor discrepancy in the functions $u$ and $v$ ($a$ and $b$ in their notation) between the original form by Ref.~\cite{MN92} and the one used by Ref.~\cite{Gair08}. The difference, however, is small, and both versions satisfy the vacuum Einstein equations. I will use the version quoted by Gair et al.~\cite{Gair08} in this paper. Following Ref.~\cite{Gair08}, I define
\ba
\alpha &\equiv& \frac{-M+\sqrt{M^2-a^2}}{a}, \nonumber \\
k &\equiv& M \frac{1-\alpha^2}{1+\alpha^2} = \sqrt{M^2-a^2}, \nonumber \\
\beta &\equiv& q \frac{M^3}{k^3},
\ea
where $M$ and $a$ are the mass and the spin, respectively. 

This metric can be mapped to cylindrical coordinates:
\ba
\rho &\equiv& k \sqrt{x^2-1}\sqrt{1-y^2},\qquad
z \equiv kxy
\ea
with inverse
\ba
x &=& \frac{1}{2k} \left[ \sqrt{\rho^2+(z+k)^2} + \sqrt{\rho^2+(z-k)^2} \right], \nonumber \\
y &=& \frac{1}{2k} \left[ \sqrt{\rho^2+(z+k)^2} - \sqrt{\rho^2+(z-k)^2} \right].
\ea
The MN metric then becomes
\begin{equation}
ds_{\MN}^2 = -f_{\MN}(dt-\omega d\phi)^2 + f_{\MN}^{-1} \left[ e^{2\Gamma}(d\rho^2+dz^2) + \rho^2d\phi^2 \right].
\end{equation}
In spherical coordinates, I define~\cite{MN92}
\be
r \equiv kx+M, \qquad
\cos\theta \equiv y,
\ee
and the MN metric is as given in Eq.~\eqref{MN-sph-metric}.

\section{Other forms of the MGBK metric}

Certain choices of the free parameters of the MGBK metric allow for a mapping to other alternative theories of gravity. See Ref.~\cite{Vigeland:2011ji} for a detailed discussion. Here, I calculate the event horizons of the black hole metrics in Einstein-Dilaton-Gauss-Bonnet and Chern-Simons gravity using the condition in Eq.~(\ref{horeqpert1}).

\subsection{Linearized Kerr}

First, however, I consider the Kerr metric, linearized in $a^2$, which can be regarded as a ``deformed Schwarzschild" spacetime. Expanding the metric element
\be
g_{rr}^{\rm K} = \frac{\Sigma}{\Delta}
\ee
to $\mathcal{O}(a^2)$ and defining $h_{rr}^{\rm dS}$ by subtracting the corresponding element of the Schwarzschild spacetime,
\be
g_{rr}^{\rm Schw} = \frac{1}{f_S(r)},
\ee
where
\be
f_S(r) \equiv 1-\frac{2M}{r},
\ee
I find
\be
h_{rr}^{\rm dS} = \frac{a^2}{r^2 f_S(r)} [ \cos^2\theta - f_S(r)^{-1} ].
\ee

Equation~(\ref{horeqpert1}) becomes
\be
f_S(r)^2 h_{rr}^{\rm dS} = \frac{a^2}{r^2} [ f_S(r)\cos^2\theta - 1 ] = 0.
\label{dSsol1}
\ee
When solving this equation, I need to bear in mind that my analysis is done to $\mathcal{O}(a^2)$, and so the horizon will be shifted by this amount
[recall Eq.~(\ref{perturbedhor})]. I insert
\be
r_H = 2M\left( 1 + \lambda \frac{a^2}{M^2} \right)
\ee
into Eq.~(\ref{dSsol1}), linearize in $a^2$, and find
\be
2Ma^2(1+4\lambda) = 0.
\ee
Thus, $\lambda=-1/4$, and the horizon of my linearized Kerr spacetime is at
\be
r_H = 2M \left( 1 - \frac{1}{4} \frac{a^2}{M^2} \right) = 2M-\frac{1}{2}\frac{a^2}{M}.
\ee
This is of course is just the exact solution for the Kerr horizon, $r_+ = M+\sqrt{M^2-a^2}$, expanded to $\mathcal{O}(a^2)$.

\subsection{Einstein-dilaton-Gauss-Bonnet gravity}

Yunes and Stein \cite{Yunes:2011we} looked at black holes in a class of gravity theories described by Lagrangians modified from
the standard Einstein-Hilbert form by scalar fields coupled to quadratic curvature invariants. They find a set of
such theories that admit black hole solutions, though deformed from the usual Kerr form. Focusing on nonrotating
solutions, they find that for a class of such solutions the relevant component of the metric deformation is given by
\ba
&& h_{rr}^{\rm EDGB} =  - \frac{\alpha_3}{\kappa M^2 r^2 f_S(r)^2}\nn \\
&& \left( 1 + \frac{M}{r} + \frac{52}{3}\frac{M^2}{r^2} + \frac{2M^3}{r^3} + \frac{16}{5}\frac{M^4}{r^4} - \frac{368}{3}\frac{M^5}{r^5} \right).
\ea
The constant $\alpha_3$ is the theory's coupling constant, and $\kappa = 1/(16\pi)$. Note that this spacetime
can also be regarded as a deformed Schwarzschild case, with a form for $\bar{\gamma}_1(r)$ that is somewhat different than
that used in Ref.~\cite{Vigeland:2011ji}.

I use this $h_{rr}^{\rm EDGB}$ in Eq.~(\ref{horeqpert1}), and use $g^{\rm Schw}_{rr} = f_S(r)$. My solution should now take the form
\be
r_H = 2M (1 + \lambda \alpha_3) .
\ee
Inserting this expression into Eq.~(\ref{horeqpert1}) and linearizing in $\alpha_3$, I find
\be
\lambda + \frac{49}{80} \frac{\alpha_3}{M^4 \kappa} = 0,
\ee
and thus
\be
r_H = 2M \left( 1 - \frac{49}{80}\frac{\alpha_3}{M^4 \kappa} \right).
\ee
This expression agrees with the text following Eq.~(12) of Ref.~\cite{Yunes:2011we}.

\subsection{Chern-Simons gravity}

As a final example, I consider slowly rotating black holes in dynamical Chern-Simons gravity, a case analyzed by
Yagi, Yunes, and Tanaka \cite{YYT12}. The relevant metric component in this case is given by
\ba
&& h_{rr}^{\rm CS} = \frac{a^2}{r^2 f_S(r)} [\cos^2\theta-f_S^{-1}(r) ] + \zeta \chi^2 \frac{M^3}{r^3 f_S^2(r)} \nn \\
&& \times \left[ \frac{201}{896} f_S(r) n_1(r) P_2(\cos\theta) - \frac{25}{384} \frac{M}{r} n_2(r) \right].
\ea
Here, $\zeta\chi^2$ are a combination of coupling parameters that determine the strength of the Chern-Simons modification
to gravity to this order, $P_2 = (3 \cos^2 \theta - 1)/2$ is a Legendre polynomial, and $n_{1,2}(r)$ are polynomials in $(M/r)$, which are given in Ref.~\cite{Yunes:2011we}.

In this case, I expect the horizon to be at
\be
r_H = 2M \left( 1 + \lambda_1 \frac{a^2}{M^2} + \lambda_2 \zeta\chi^2 \right).
\ee
Building the horizon equation~(\ref{horeqpert1}), inserting this form for $r_H$, and linearizing in $a^2$ and $\zeta\chi^2$, I find
\be
\frac{a^2}{M^2} \left( \frac{1}{4} + \lambda_1 \right) + \zeta\chi^2 \left( \frac{915}{57344} + \lambda_2 \right) = 0.
\ee
This means
\ba
r_H &=& 2M \left( 1 - \frac{1}{4} \frac{a^2}{M^2} - \zeta\chi^2 \frac{915}{57344} \right) \nn \\
&=& 2M - \frac{1}{2} \frac{a^2}{M} - \frac{915}{28672}\zeta\chi^2 M.
\ea
This agrees with Eq.~(55) of Ref.~\cite{YYT12}, linearized in $a^2$.

\section{Numerical methods for finding event horizons}

Here I describe two numerical methods for solving Eq.~(\ref{hor_master}) in order to construct the event horizon.

\subsection{Spectral method}

The following technique only works if the horizon's geometry can be described as a ``Strahlk\"orper" \cite{Thornburg}. A
Strahlk\"orper is a figure that encloses the origin and is only intersected once by any ray from the origin. This is
the case for mild deformations from Kerr, but may not be the case if the deformation is more severe. A horizon of unusual topology (e.g., disjoint horizons on the $z$ axis, or a torus) is certainly not a Strahlk\"orper, at least not about ``r = 0". It may be possible to adapt this technique to such situations by changing the origin (e.g., to the middle of a disjoint horizon, or to the center ring of a torus).

Bearing in mind my boundary conditions [$H(0) = H_K$, $H(\pi/2) = H_{\rm eq}$, $dH/d\theta = 0$ at $\theta = 0,\pi/2$], I proceed by
writing
\be
H(\theta) = \sum_{n=0}^N \alpha_n P_n(\cos\theta),
\ee
where $P_n$ are the Legendre polynomials and where $N$ is a suitably chosen truncation of the sum over them. Solving for $H(\theta)$ then means solving for the expansion coefficients $\alpha_n$.

All of the spacetimes that I consider are reflection symmetric about $\theta = \pi/2$, so I need only include even values
of $n$. I then modify the expansion to
\be
H(\theta) = \sum_{n=0}^N \alpha_{2n} P_{2n}(\cos\theta).
\ee
I choose $N-1$ angles evenly spaced between $\theta = 0$ and $\theta = \pi/2$. By enforcing Eq.~(\ref{hor_master}) at these $N-1$ angles,
plus the solutions at $\theta = 0$ and $\theta = \pi/2$ from Eq.~(\ref{grrzero}), I have a total of $N + 1$ equations for the $N + 1$ coefficients, completely
specifying the horizon function $H(\theta)$.

\subsection{Finite difference method}

A crude but robust method for solving the horizon equation is to use finite difference to construct the derivative
and solve the equation. Cover the angular sector with $N$ segments of width $\delta\theta$, so that there are a total of $N + 1$
angles from $\theta_0 = 0$ to $\theta_N = \pi/2$. Let $H_i$ denote the solution at angle $\theta_i$. Then, the horizon equation~(\ref{hor_master}) becomes
\be
0 = g^{rr}(H_i, \theta_i) + g^{\theta\theta}(H_i, \theta_i) \left( \frac{ H_{i-1}-H_i }{ \delta\theta } \right)^2.
\ee
The solution is known at $i = 0$ ($\theta = 0$); from there, step to $i = N$ ($\theta = \pi/2$). Empirically, I have found that
this method agrees with the spectral method extremely well if there is a null surface that encloses the origin and passes
through $\theta = 0$ and $\theta = \pi/2$.

It may be possible to generalize this method to handle different null surface topologies. Since I have not identified any parameter choices that unambiguously yield such unusual null surfaces, I have not tested this. Note that the null surfaces in the MK metric, where they exist, always have spherical topology, while the Killing horizon can have spherical, disjoint, or toroidal topology.


\end{document}